\documentclass[twocolumn]{aastex63}

\usepackage{apjfonts}

\def\d3{$\delta_{3}$ }
\def\1d3{$(1 + \delta_{3})$ }
\def\l1d3{$\log_{10}(1 + \delta_{3})$ }

\def\s3{$\Sigma_{3}$}
\def\ha{H$\alpha$}
\def\hb{H$\beta$}

\def\othree{[OIII] 5007}
\def\ntwo{[NII] 6584}

\def\24m{24 $\mu$m}

\def\sm{$\rm~M_{*}$}

\def\kms{${\rm km~s^{-1}}$ }

\def\Msolar{$\rm M_{\odot}$}

\def\rmxaa{RMxAA}

\def\sigsm{$\Sigma_{*}$}
\def\sigsfr{$\Sigma_{\rm SFR}$}

\def\sigdense{$\Sigma_{\rm dense}$}

\def\h2{$\rm H_{2}$}
\def\Mdense{$\rm M_{dense}$}
\def\Mmol{$\rm M_{mol}$}

\def\Mh2{$\rm M_{H_{2}}$}

\def\sigh2{$\Sigma_{\rm H_{2}}$}
\def\sigmol{$\Sigma_{\rm mol}$}
\def\fgas{$f_{\rm gas}$}
\def\fh2{$f_{\rm H_{2}}$}
\def\fmol{$f_{\rm mol}$}

\def\fdense{$f_{\rm dense}$}
\def\Rdense{$R_{\rm dense}$}

\def\Re{$R_{e}$}
\def\co{$^{12}$CO(1-0)}

\def\HCO{$\rm HCO^{+}$}
\def\SFEmol{$\rm SFE_{\rm mol}$}
\def\SFEdense{$\rm SFE_{\rm dense}$}

\shorttitle{Dense Molecular Gas in Green Valley Galaxies}
\shortauthors{Lin et al.}

\begin{document}

\title{The ALMaQUEST Survey XII: Dense Molecular Gas as traced by HCN and \HCO~in Green Valley Galaxies}

\author{Lihwai Lin}
\altaffiliation{Email: lihwailin@asiaa.sinica.edu.tw}
\affiliation{Institute of Astronomy \& Astrophysics, Academia Sinica, Taipei 10617, Taiwan}

\author[0000-0002-1370-6964]{Hsi-An Pan}
\affiliation{Department of Physics, Tamkang University, No.151, Yingzhuan Road, Tamsui District, New Taipei City 251301, Taiwan} 

\author{Sara L. Ellison}
\affiliation{Department of Physics \& Astronomy, University of Victoria, Finnerty Road, Victoria, British Columbia, V8P 1A1, Canada}

\author{Nanase Harada}
\affiliation{National Astronomical Observatory of Japan, 2-21-1 Osawa, Mitaka, Tokyo 181-8588, Japan}
\affiliation{Astronomical Science Program, Graduate Institute for Advanced Studies (SOKENDAI), 2-21-1 Osawa, Mitaka, Tokyo, 181-1855 Japan}

\author{Mar\'{i}a J. Jim\'{e}nez-Donaire}
\affiliation{Observatorio Astron\'{o}mico Nacional (IGN), C/Alfonso XII, 3, 28014 Madrid, Spain}
\affiliation{Centro de Desarrollos Tecnol\'{o}gicos, Observatorio de Yebes (IGN), 19141 Yebes, Guadalajara, Spain}

\author[0000-0002-4235-7337]{K. Decker French}
\affiliation{Department of Astronomy, University of Illinois, 1002 W. Green Street, Urbana, IL 61801, USA}

\author{William M. Baker}
\affiliation{Cavendish Laboratory, University of Cambridge, 19 J. J. Thomson Avenue, Cambridge CB3 0HE, United Kingdom}
\affiliation{Kavli Institute for Cosmology, University of Cambridge, Madingley Road, Cambridge, CB3 OHA,UK.}

\author{Bau-Ching Hsieh}
\affiliation{Institute of Astronomy \& Astrophysics, Academia Sinica, Taipei 10617, Taiwan}

\author{Yusei Koyama}
\affiliation{Subaru Telescope, National Astronomical Observatory of Japan, 650 North A\'ohoku Place, Hilo, HI 96720,USA}
\affiliation{Astronomical Science Program, Graduate Institute for Advanced Studies (SOKENDAI), 2-21-1 Osawa, Mitaka, Tokyo, 181-1855 Japan}

\author{Carlos L\'{o}pez-Cob\'{a}}
\affiliation{Institute of Astronomy \& Astrophysics, Academia Sinica, Taipei 10617, Taiwan}

\author{Tomonari Michiyama}
\affiliation{Faculty of Welfare and Information, Shunan University, 43-4-2 Gakuendai, Shunan, Yamaguchi, 745-8566, Japan}

\author{Kate Rowlands}
\affiliation{Space Telescope Science Institute, 3700 San Martin Dr Baltimore, MD 21218, USA}

\author{Sebasti\'{a}n F. S\'{a}nchez }
\affiliation{Instituto de Astronom\'ia, Universidad Nacional Aut\'onoma de  M\'exico, Circuito Exterior, Ciudad Universitaria, Ciudad de M\'exico 04510, Mexico}

\author{Mallory D. Thorp}
\affiliation{Argelander-Institut f\"{u}r Astronomie, Universit\"{a}t Bonn, Auf dem H\"{u}gel 71, 53121 Bonn, Germany}

\begin{abstract}
 
We present ALMA observations of two dense gas tracers, HCN(1-0) and HCO$^{+}$(1-0), for three galaxies in the green valley and two galaxies on the star-forming main sequence with comparable molecular gas fractions as traced by the CO(1-0) emissions, selected from the ALMaQUEST survey. We investigate whether the deficit of molecular gas star formation efficiency ($\rm SFE_{\rm mol}$) that leads to the low specific star formation rate in these green valley galaxies is due to a lack of dense gas (characterized by the dense gas fraction $f_{\rm dense}$) or the low star formation efficiency of dense gas ($\rm SFE_{\rm dense}$). We find that $\rm SFE_{\rm mol}$ as traced by the CO emissions, when considering both star-forming and retired spaxels together, is tightly correlated with $\rm SFE_{\rm dense}$ and depends only weakly on $f_{\rm dense}$. The specific star formation rate (sSFR) on kpc scales is primarily driven by \SFEmol~and \SFEdense, followed by the dependence on $f_{\rm mol}$, and is least correlated with $f_{\rm dense}$ or the dense-to-stellar mass ratio ($R_{\rm dense}$). When compared with other works in the literature, we find that our green valley sample shows lower global $\rm SFE_{\rm mol}$ as well as lower $\rm SFE_{\rm dense}$ while exhibiting similar dense gas fractions when compared to star-forming and starburst galaxies.
We conclude that the star formation of the 3 green valley galaxies with a normal abundance of molecular gas is suppressed mainly due to the reduced $\rm SFE_{\rm dense}$ rather than the lack of dense gas.

\end{abstract}


\section{INTRODUCTION}
The link between molecular gas, which is the fuel of star formation, and the rate of forming stars, has been conventionally characterized using the Schmidt-Kennicutt relation \citep[hereafter SK relation;][]{sch59,ken98}. The SK relation connects the star formation rate (SFR) surface density (\sigsfr) to the molecular gas surface density (\sigmol) as \sigsfr~ $\propto$ \sigmol$^{n}$ (or equivalently, Log \sigsfr~= $n \times$Log \sigmol + constant). The power index $n$ is found close to unity but varying between 0.7 and 1.4, depending on the galaxy types \citep[late vs. early types; main sequence vs. green valley vs. starburst galaxies; AGN vs. non-AGN hosts, e.g.,][]{col18,san18,ken21,lin22,bak23}, physical scales \citep{lin19b,ell21a,pes21,bak22}, galactic environments \citep{pes22}, large-scale environments \citep{jim23}, local conditions of star formation \citep[star-forming vs. retired regions;][]{ell21b,lin22}, and gas tracers \citep{won02,gao04,big08}.

Among those investigations, the exploration of the molecular gas abundance and its properties at (sub)-kpc scales across different galaxy populations has received significant attention due to the emergence of extensive large optical integral field unit (IFU) surveys and spatial resolution matched observations facilitated by ALMA. Recent studies on kpc scales reveal that the star formation efficiency (\SFEmol~ = \sigsfr/\sigmol) measured on kpc scales depends on global galaxy properties, such as the global specific star formation rate \citep[sSFR;][]{sch11,ler13,uto18,col18,lin19b,bro20,sun20,ell21a,lin22,bak23}. For example, based on CO(1-0) observations from the ALMA-MaNGA QUEnching and STar formation \citep[ALMaQUEST;][]{lin20} survey, \citet{lin22} found that green valley (hereafter GV) galaxies not only show lower molecular gas fractions (\fmol), defined as the ratio of the molecular gas surface density to the stellar mass surface density (\sigmol/\sigsm), but also depart from the main sequence (hereafter MS) galaxies in the resolved SK relation (rSK) toward a lower value of \sigsfr~ at a given \sigmol~ \citep[also see][]{lin17,bro20,ell21a}. In other words, green valley galaxies exhibit lower local SFE compared to the galaxies located on the star-forming main sequence, similar to the findings based on global studies \citep[e.g.,][]{sai17}.  However, the physical process that controls SFE remains unclear. 

While the CO emission lines, in particular the low-$J$ transition such as CO(1-0), are commonly used as the standard tracer of the bulk of molecular gas, the critical density of CO(1-0) emission line is on the order of $\sim$10$^{3}$cm$^{-3}$ and therefore traces not only dense but also diffuse gas that is not necessarily capable of forming stars. On the other hand, HCN and \HCO~ have a greater critical density ($>$ 10$^{4}$cm$^{-3}$) and are suggested to better correlate with the star formation rate of galaxies \citep{gao04,gar12,use15,jim19,neu23}. The molecular gas SFE$_{\rm mol}$ can be written as:

\begin{eqnarray}\label{eq:sfe}
\rm SFE_{\rm mol} &=& \rm SFR/\rm M_{\rm mol}\nonumber\\
&=& \rm SFR/\rm M_{\rm dense} \times \rm M_{\rm dense}/\rm M_{\rm mol}\nonumber\\
&=&   \rm SFE_{\rm dense} \times f_{\rm dense},
\end{eqnarray}
where \Mmol~and \Mdense~are the total molecular gas mass and dense molecular gas mass, respectively, SFE$_{\rm dense}$ is the dense gas star formation efficiency, and $f_{\rm dense}$ refers to the dense-to-molecular gas ratio, i.e., the dense gas fraction. Therefore, it leaves two possible explanations for the lower $\rm SFE_{mol}$ of green valley galaxies relative to that of main-sequence galaxies. The low SFE$_{\rm mol}$ of green valley galaxies can be attributed to either lower \SFEdense~ or lower \fdense~ than that in main sequence galaxies. Since dense gas is normally more spatially concentrated than both the CO(1-0) and the stellar components \citep{jia20}, using the global HCN/CO ratio does not properly capture the variations in gas phase quantities. Spatially resolved observations are therefore critical for assessing star formation tracers, CO, and HCN (and \HCO) on well-matched scales.

A number of previous works have investigated the role of dense gas in star formation on spatially resolved scales.
For example, \citet{lon13} found that \SFEdense~ in Galactic Center of our own Milky Way is significant lower than in the rest of the disk. Detailed  studies of individual nearby galaxies revealed that SFE can also depend on the dynamical environments of galaxies \citep[e.g.,][]{que19,san22}. In addition to these case studies, dense gas observations over a statistical sample allow for quantitative characterization of the dependence of \SFEdense~on various galaxies properties. The EMIR Multiline Probe of the ISM Regulating Galaxy Evolution \citep[EMPIRE; PI: F. Bigiel; ][]{jim19} survey observed 9 nearby disk galaxies with HCN(1-0), \HCO(1-0), and HNC(1-0) lines. The MALATANG \citep[Mapping the dense molecular gas in the strongest star-forming galaxies;][]{tan18,jia20} survey mapped the HCN(4-3) and \HCO(4-3) line emissions in 23 of the nearest, IR-brightest galaxies with the JCMT telescope.  Based on the study of IR-to-HCN measurements of 29 disk galaxies, \citet{use15} found that the inferred \SFEdense~ show a strong dependence on the location in a disk assuming a fixed conversion factor. The ACA Large-sample Mapping Of Nearby galaxies in Dense gas \citep [ALMOND;][]{neu23} studied the dependence of kpc-scale (2.1 kpc) HCN/CO and SFR/HCN ratios on various structural and dynamical properties of the cloud-scale (150 pc) molecular gas across 25 nearby galaxies. Nevertheless, existing surveys predominantly concentrate on star-forming and IR-bright sources, and spatially resolved dense gas observations for objects with depleted star formation remain scarce. Most recently, \citet{fre23} reported dense gas HCN(1-0), \HCO(1-0), and HNC (1-0) observations for six post-starburst (PSB) galaxies and found that most PSB galaxies have low HCN/CO and \HCO/CO line ratios ($< 0.04$), suggesting that the low SFR of the PSB population is likely due to the lack of dense gas.

In this work, we obtain ALMA observations for HCN and \HCO~emissions as dense tracers with the spatial resolution matched to the existing CO(1-0) observations for 3 GV and 2 MS galaxies selected from the ALMaQUEST survey \citep{lin20}. We focus on distinguishing the effects between \fdense~and \SFEdense~on driving \SFEmol.  As the HCN and \HCO~ emissions are in general fainter than CO by more than an order of magnitude, we start with the CO-bright sample for both GV and MS targets, allowing us to obtain a possible HCN and \HCO~ detection within a reasonable amount of ALMA integration time. The sample is thus biased toward galaxies that are not deficient in molecular gas. However, this sample still offers opportunities to investigate the central question we attempt to answer: Is the SFE$_{\rm mol}$ primarily regulated by the fraction of molecular gas in the dense phase (\fdense) or \SFEdense?  Across this paper, the subscript `mol' is adopted to refer to the molecular gas quantities obtained from CO(1-0) observations, whereas the subscript `dense' is used to refer to the dense gas quantities based on HCN(1-0) or \HCO (1-0) observations.  

In \S2, we describe the sample selections and the observational data used in this work. In \S3, we present the resolved relation between SFR surface density and dense gas mass surface density (the dense gas version of Schmidt-Kennicutt relation) and investigate the dependence of SFE$_{\rm mol}$ on the SFE$_{\rm dense}$ and the dense-to-molecular ratio (\fdense). In \S4, we discuss issues affecting the interpretation of this work. Conclusions are given in \S5.

Throughout this paper we adopt the following cosmology: \textit{H}$_0$ = 70~\kms Mpc$^{-1}$, $\Omega_{\rm m} = 0.3$ and $\Omega_{\Lambda } = 0.7$. We use a Salpeter initial mass function (IMF).

\begin{deluxetable*}{llccccc}
\tabletypesize{\scriptsize}
\tablewidth{0pt}
\tablecaption{Global Properties of HCN and \HCO~targets \label{tab:sample}}
\tablehead{
    \colhead{Plate-IFU} &
    \colhead{Galaxy Type} &
    \colhead{RA} &
    \colhead{DEC} &
    \colhead{MaNGA redshift} &
    \colhead{log$_{10}$(\sm/\Msolar)$^{(a)}$} &
    \colhead{log$_{10}$($\frac{\rm SFR}{\rm M_{\odot} yr^{-1}}$)$^{(b)}$}\\
&&(deg)&(deg)}
\startdata
7815-12705 & MS & 318.990448 & 9.543076 & 0.029550 & 10.83 & 0.42 \\
 8081-9101 & MS & 332.798737 & 11.800733 & 0.028460 & 10.60 & 0.32 \\
 8081-12703 & GV & 331.122894 & 12.442626 & 0.025583 & 10.34 & -0.79 \\
 8083-6101 & GV & 332.892853 & 11.795929 & 0.026766 & 10.30 & -0.88 \\
 8950-12705 & GV & 42.032784 & -0.752316 & 0.025277 & 10.53 & -0.41
\enddata
\tablecomments{$^{(a)}$ $^{(b)}$ Values are taken from value-added catalog \citep{san18} in the SDSS DR15 release \citep{aqu19}.}
\end{deluxetable*}

\begin{deluxetable*}{lcccccccccc}
\tabletypesize{\scriptsize}
\tablewidth{0pt}
\tablecaption{Global Properties of HCN and \HCO~targets enclosed by 1.5 \Re\label{tab:obs}}
\tablehead{
    \colhead{Plate-IFU} &
    \colhead{log$_{10}$($\frac{\rm SFR}{\rm M_{\odot} yr^{-1}}$)$^{(a)}$} &
    \colhead{$\sigma_{\rm HCN}^{(b)}$} &
    \colhead{$\sigma_{\rm HCO^{+}}^{(c)}$} &
    \colhead{I$_{\rm CO}$}$^{(d)}$ &
    \colhead{I$_{\rm HCN}$}$^{(e)}$ &
    \colhead{I$_{\rm HCO^{+}}$}$^{(f)}$ &
    \colhead{log$_{10}$(L$_{\rm CO}$/L$_{\odot}$)}$^{(g)}$ &
    \colhead{log$_{10}$(L$_{\rm HCN}$)/L$_{\odot}$}$^{(h)}$ &
    \colhead{log$_{10}$(L$_{\rm HCO^{+}}$)$^{(i)}$/L$_{\odot}$}\\
&&(Jy beam$^{-1}$ \kms)&(Jy beam$^{-1}$ \kms)&(Jy \kms)&(Jy \kms)&(Jy \kms)}
\startdata
7815-12705 & 0.47 & 0.0148 & 0.0154 & 32.27729 & 0.74242 & 0.85672 & 8.75 & 7.34 & 7.40 \\
 8081-9101 & 0.24 & 0.0224 & 0.0174 & 18.95202 & 0.54407 & 0.51363 & 8.65 & 7.34 & 7.31 \\
 8081-12703 & -1.23 & 0.0135 & 0.0167 & 7.75231 & 0.23821 & 0.15411 & 7.84 & 6.55 & 6.36 \\
 8083-6101 & -1.12 & 0.0139 & 0.0112 & 17.38613 & 0.55128 & 0.42021 & 8.25 & 6.98 & 6.86 \\
 8950-12705 & -1.10 & 0.0227 & 0.0201 & 24.11672 & 0.30664 & 0.75681 & 8.04 & 6.37 & 6.76
\enddata
\tablecomments{$^{(a)}$ The total star formation rate converted from \ha~ emissions integrated over 1.5 \Re. $^{(b)}$ $^{(c)}$ The 1$\sigma$ sensitivity of the integrated ALMA intensity maps, calculated using the spectral window shown as the shaded areas in Figure \ref{fig:spec_gas}. $^{(d)}$ $^{(e)}$ $^{(f)}$ The total line intensity integrated over 1.5 \Re. $^{(g)}$ $^{(h)}$ $^{(i)}$ The total line luminosity integrated over 1.5 \Re. }
\end{deluxetable*}

\begin{figure*}[tbh]

\includegraphics[scale=0.45]{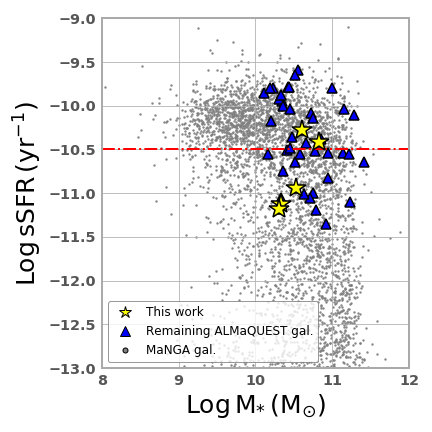}
\includegraphics[scale=0.3]{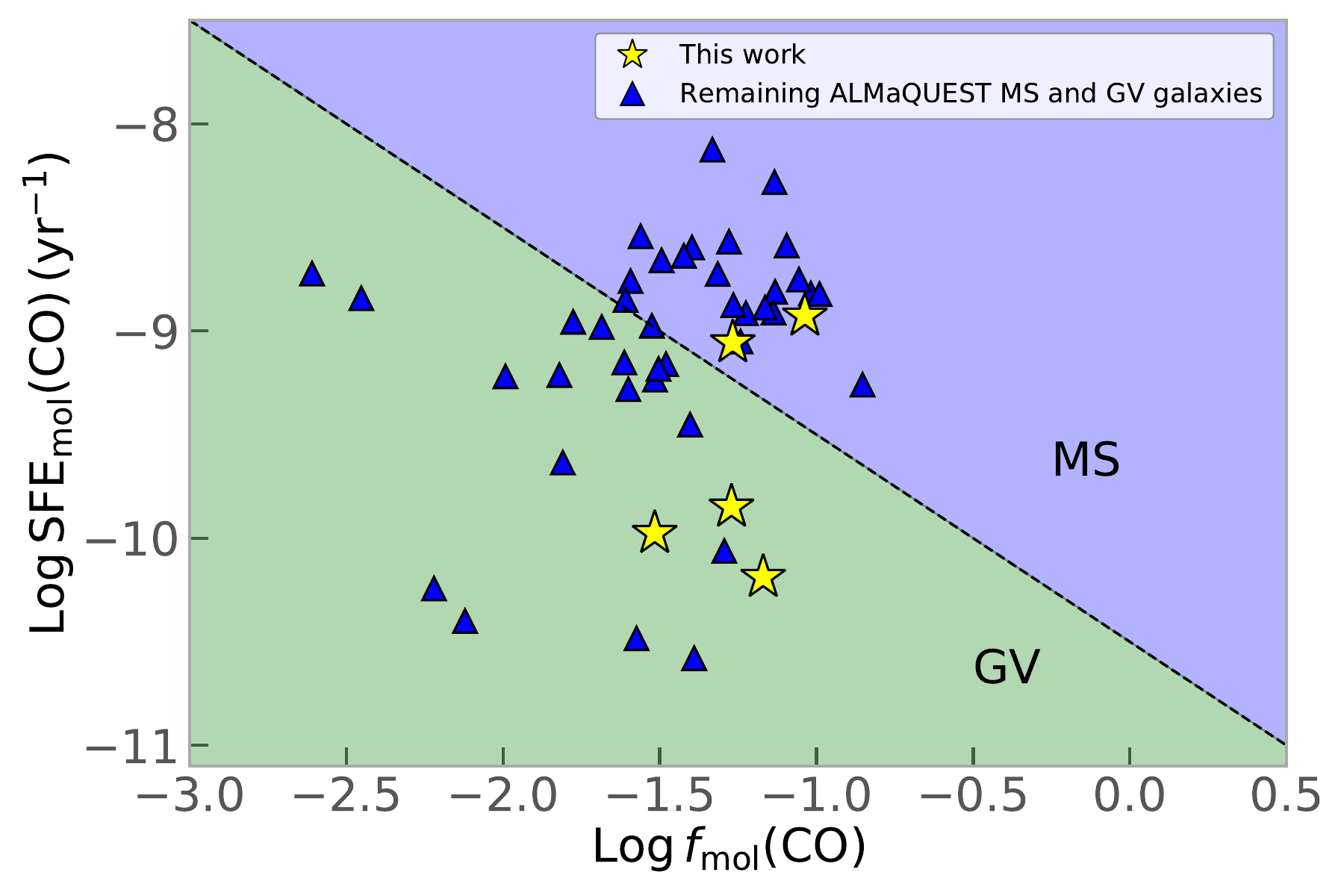}
\caption{Left panel: The global specific star formation rate (sSFR) vs. stellar mass (\sm) of 4656 MaNGA galaxies (black dots) from the Pipe3D \citep{san16a,san16b} value-added catalog \citep{san18} in the SDSS DR15 release \citep{aqu19}. The yellow star symbols denote the 5 ALMaQUEST galaxies for which we observed in ALMA HCN/\HCO~used in this work. The remaining 41 ALMaQUEST galaxies are shown in blue triangles. The red lines denotes the dividing line defining the main sequience (MS) and green valley (GV) subsamples. Our HCN (and \HCO) targets are among those with the lowest sSFR rate in the ALMaQUEST green valley sample. Right panel: The molecular gas star formation efficiency (\SFEmol) vs. molecular gas fraction (\fmol) computed within 1.5 \Re~using CO for the ALMaQUEST MS and GV subsamples \citep{lin22}. The green and blue shaded areas denote the regions that occupy the main sequence and green valley galaxies, respectively, and the color division corresponds to the red dividing line (Log$_{10}$sSFR = -10.5) in the left panel. The color symbols are the same as those in the left panel. The three green valley targets with HCN observations have similar CO-based gas fractions but much lower star formation efficiency compared to the main sequence targets. \label{fig:property}}
\end{figure*}

\begin{figure*}
\centering
\includegraphics[angle=0,width=0.95\textwidth]{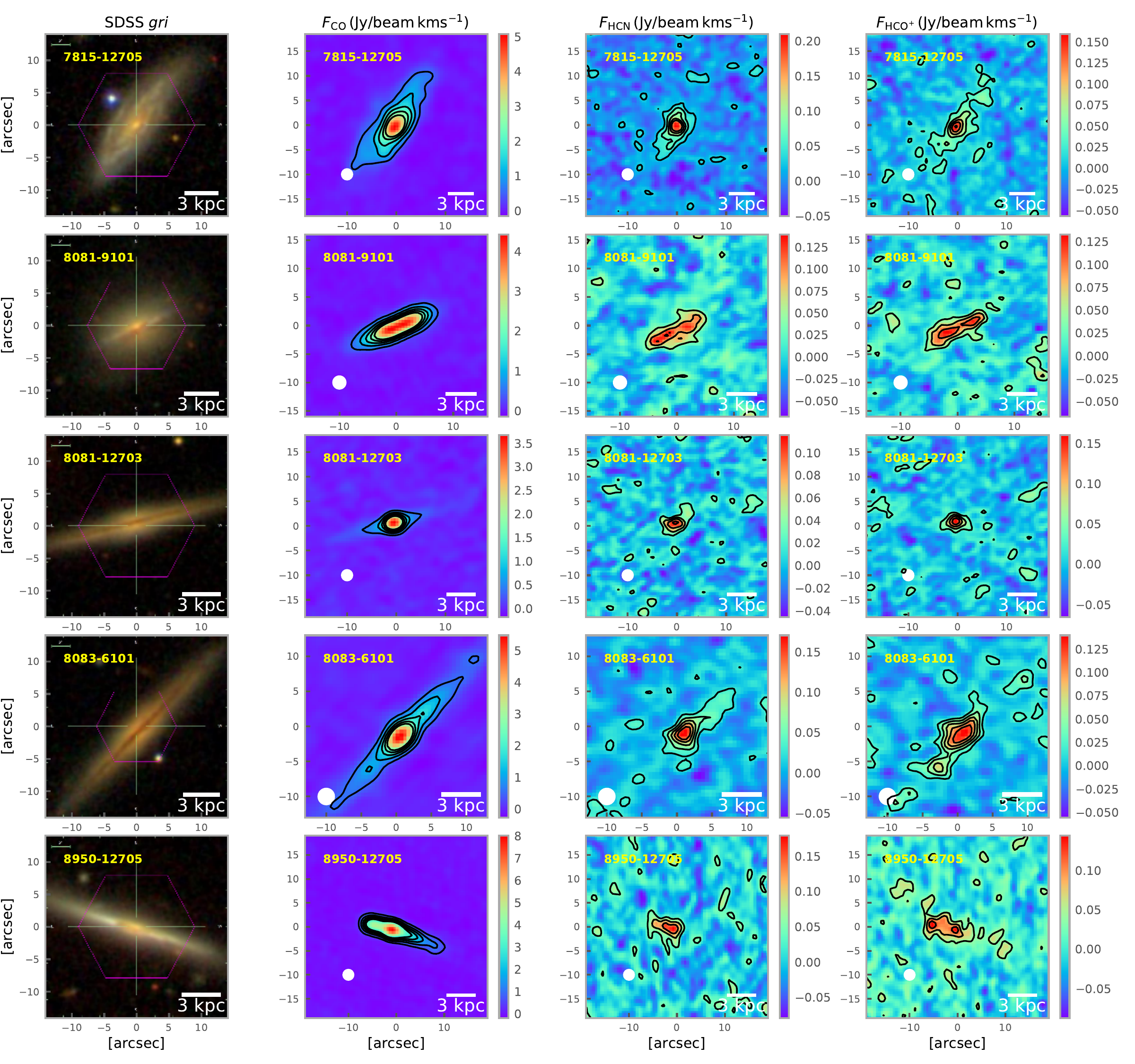}
\caption{ALMA observations for 5 ALMaQUEST galaxies used in this work. The top two rows are for main sequence galaxies and the bottom 3 rows are for green valley galaxies. From left to right: SDSS $gri$ composite images, \co~intensity  (Jy km s$^{-1}$ per beam), HCN(1-0) intensity  (Jy km s$^{-1}$ per beam), and  \HCO(1-0) intensity  (Jy km s$^{-1}$ per beam) maps. The back contours denote regions with S/N = (2,4,6,8,10) for HCN and \HCO, and regions with S/N = (10,20,30,40,50) for CO observations. The white circle in the lower-left corner of each gas maps illustrates the restoring beamsize. The magenta hexagons denote the MaNGA footprints.  \label{fig:map_gas}}
\end{figure*}

\begin{figure}
\centering
\includegraphics[angle=0,width=0.47\textwidth]{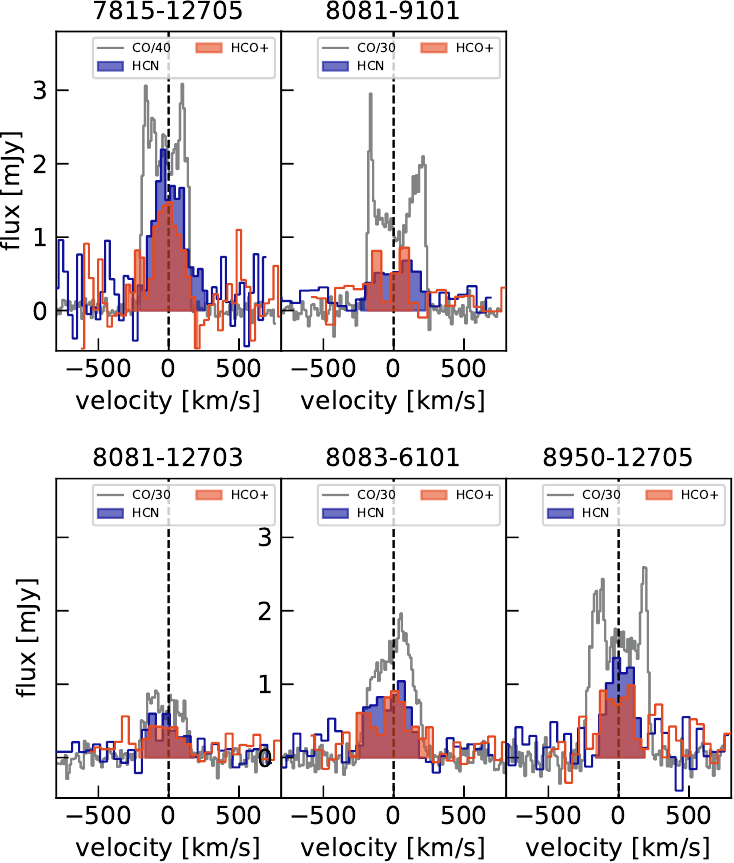}

\caption{Integrated spectra over the area enclosed by the 1.5 \Re~ with a binned spectral resolution ranging from 30-50 \kms~for the 5 targets presented in this work. The MaNGA plate-IFU identifier is given above each panel. The blue and red shaded areas represent the region of the spectrum used for computing the HCN and \HCO~line fluxes, respectively. For comparison, the CO spectra binned at a spectral resolution of 11 \kms~are overplotted and the CO fluxes have been adjusted by applying a scaling factor to fit the figure. \label{fig:spec_gas}}
\end{figure}

\section{SAMPLE and OBSERVATIONS \label{sec:data}}
In this work, we study 5 galaxies selected from the ALMaQUEST sample \citep{lin20}, which observed 46 galaxies chosen from the Mapping Nearby Galaxies at Apache Point Observatory \citep[MaNGA;][]{bun15} survey with the Atacama Large Millimeter Array (ALMA) of CO(1-0). Details on the CO(1-0) data reduction and properties  are described in the main survey paper \citep{lin20} and the global properties of the molecular gas in ALMaQUEST green valley galaxies are presented in \citet{lin22}. Table \ref{tab:sample} lists the general information of the 5 targets analyzed in this work. Dense gas observations in HCN(1-0) (restframe 88.631 GHz) and \HCO(1-0)(restframe 89.189 GHz) of the 5 ALMaQUEST galaxies were carried out with ALMA in Cycle 7 on 2019 using the Band 3 receiver and C43-2 configuration (project code: 2019.1.01178.S; PI: Lihwai Lin). The synthesized beam FWHM is $\sim$ 2.5$\arcsec$ that is matched to the MaNGA and CO (1-0) spatial resolutions, which correspond to 1.2-1.5 kpc for the 5 galaxies presented in this work. While the resolution is not sufficient to resolve the giant molecular clouds (GMCs), which is typically several tens of pc, the observations offer investigations at scales that bridge the global to the local properties. For the remainder of the paper, we refer to CO(1-0), HCN(1-0) and \HCO(1-0) emissions as CO, HCN and \HCO, respectively.

In Figure \ref{fig:property}, we display the global properties of the 5 targets (yellow symbols) selected for the dense gas observations with respect to the remaining 41 ALMaQUEST galaxies (blue symbols). The left panel shows the distribution of galaxies on the integrated (global) specific star formation rate (sSFR) vs. stellar mass (\sm) plane using measurements taken from the Pipe3D \citep{san16a,san16b} value-added catalog \citep{san18} in the SDSS DR15 release \citep{aqu19}. The global \fmol~vs. SFE$_{\rm mol}$ are displayed in the right panel.
We target 3 green valley galaxies with normal \fmol~ but low SFE$_{\rm mol}$ based on the CO measurements. We also observed HCN and \HCO~ for 2 main sequence galaxies, with predicted HCN and \HCO~brightest among those with similar \fmol.  While \fmol~of the 3 green valley galaxies is comparable to the 2 main sequence galaxies, their SFE$_{\rm mol}$ is suppressed by a factor of $\sim$10, making their sSFRs among the lowest ones in our sample.  This setup allows us to assess the difference in the dense gas abundance relative to the molecular gas, traced by the HCN/CO and \HCO/CO ratios, between the main sequence and green valley samples.

Our spectral setup includes one line targeting HCN and three low-resolution spectral windows for checking the continuum.
The target-line window has a bandwidth of $\sim$930 MHz (3200 km s$^{-1}$), with a native channel width of $\sim$ 2 km s$^{-1}$ and is sufficiently wide to also include the line of HCO$^{+}$.
The data were processed by the standard pipeline in the Common Astronomy Software Applications package \citep[CASA version 5.6;][]{mcm07}.
The systematic flux uncertainty associated with the calibration is typically 5\%-10\% in Band 3 (see ALMA Proposer's Guide)

The task tclean was employed for deconvolution with a robust = 0.5 weighting (Briggs). 
We adopted a user-specified image center, pixel size (0.5\arcsec), and restoring beamsize (2.5\arcsec) to match the image grid and the spatial resolution of the MaNGA images.
The restoring beamsize is similar to that of the native beamsize reported by the tclean (2.7$\arcsec$ $\times$ 2.4$\arcsec$).  
The  HCN and HCO$^{+}$ lines were imaged separately. 
To increase the signal-to-noise ratio, the spectral channels are binned to 30 to 50 km s$^{-1}$, varying from galaxy to galaxy. 
The rms noise of the spectral line data cubes ranges from $\sim$ 0.087 to 0.152 mJy/beam.

The integrated intensity maps of HCN and \HCO~ were constructed using the task IMMOMENTS in CASA. The integrated intensity maps were created by integrating emission from a velocity range set by hand to match the observed line profile without any clipping in the signal. Figure \ref{fig:map_gas} displays the maps of various emission lines, including CO, HCN, and \HCO. As can be seen, the two lines are detected in all our 5 targets. We compute the total intensities of CO, HCN, and \HCO~by summing up all the spaxel values within a 1.5 effective radius (1.5\Re) provided in the Pipe3D catalog. The total intensity and luminosity of the 5 objects considered in this work are listed in Table \ref{tab:obs}. The global HCN-to-CO and \HCO-to-CO flux ratios are measured to be 0.02-0.05 and 0.03-0.05, respectively, broadly consistent with the global values found in previous works \citep[e.g.,][]{use15,jim19}. In Figure \ref{fig:spec_gas}, we show the ALMA spectra centered on the position of the HCN and \HCO~ lines using the systematic velocity taken from the NASA Sloan Atlas (NSA) redshift\footnote{https://www.sdss4.org/dr17/manga/manga-target-selection/nsa/}, integrated over the region enclosed by 1.5 \Re. The CO spectra from the previous program \citep{lin20} are also overplotted for comparison.

In this work, the molecular gas mass surface density, \sigmol, is computed from the CO luminosity by adopting a constant conversion factor ($\alpha_{\mathrm{CO}}$) of 4.35 \Msolar~ (K km s$^{-1}$ pc$^{2}$)$^{-1}$ \citep[e.g.,][]{bol13}, which includes contributions from molecular hydrogen, helium, and heavier elements. Similarly, we convert the HCN and \HCO~luminosity into the dense gas mass by adopting the conversion factor $\alpha_{\rm HCN}$ = 10 \Msolar  (K km s$^{-1}$ pc$^{2}$)$^{-1}$ and $\alpha_{\rm HCO^{+}}$ = 10 \Msolar~  (K km s$^{-1}$ pc$^{2}$)$^{-1}$ following the suggestions given by \citet{gao04}. While $\alpha_{\mathrm{CO}}$ are found to vary from galaxy to galaxy and even within galaxies and various prescriptions have been proposed to predict $\alpha_{\mathrm{CO}}$ \citep[e.g.,][]{wol10,feld12,bol13,acc17,sun20,chi21,ten23}, the parameterization of the conversion factors of dense gas ($\alpha_{\rm HCN}$ and $\alpha_{\rm HCO^{+}}$) at kpc scales, which potentially can depend on the metallicity, remains relatively unexplored. As a result, for the purposes of this study and for making a comparison with other works in the literature, we have chosen to utilize fixed conversion factors. It is noteworthy that the spaxel-based metallicities, computed using the O3N2 diagnostic \citep{pet04,kew08}, exhibit median values ranging from 8.69 to 8.76 for each galaxy. The difference is within 0.07 dex of solar metallicity, making a fixed conversion factor for all 5 galaxies a fair assumption.

The MaNGA datacubes utilized in this work are taken from the MaNGA DR15 PIPE3D \citep{san16a,san16b} value-added products \citep{san18}, which contains both global properties of MaNGA galaxies and the spaxel-based measurements of \sigsm~ and emission-line fluxes. We apply the correction of dust extinction to emission line measurements using the Balmer decrement computed on a spaxel basis assuming an intrinsic \ha/\hb = 2.86 and a Milky Way extinction curve with Rv = 3.1 \citep{car89}. The extinction corrected \ha~flux is then converted to the SFR following the prescription given by Kennicutt (1998) with a Salpeter IMF. To compute \sigsm~ and \sigsfr, we normalize the stellar mass and SFR derived for each spaxel to the physical area of one spaxel corrected for inclination taken from NSA. As discussed in \citet{lin22}, there can be mixed contributions from both the newly formed stars and old stellar populations to the \ha~emission, where the former dominant in the star-forming spaxels whereas the latter dominates the retired spaxels \citep[e.g.,][]{sta08,sin13,smi22}. Following \citet{lin22}, we assume that the \ha~emission in the star-forming spaxels is purely from active star formation. On the other hand, since the \ha~emission in the retired regions is mostly powered by evolved stars rather than new star formation, the SFR estimated using the \ha~to SFR conversion serves only as an upper limit for retired spaxels \citep{sar10,yan12,sin13,hsi17,bel17,can19,ell21b}.

We limit our analyses to spaxels having signal-to-noise (S/N) > 3 for strong emission lines, i.e., \ha~and~\hb~ lines, S/N > 2 for weak lines, such as the [OIII] and [NII] lines, and S/N > 2.5 for CO, HCN, and \HCO~lines. The choice of the S/N threshold is made by balancing the uncertainty and the number of spaxels. The trends presented in this work are found to be stable against the lower limits of the S/N cut. On the other hand, it is worth noting that the upper limit of the S/N ratio does have an impact on the comparison of the correlation coefficients between different gas tracers when investigating the Schmidt-Kennicutt relations presented in \S \ref{sec:sk_dense}. We will return to this point in \S \ref{sec:discussion}.

Similar to the methodology we adopted in \citet{lin22}, we first classify each spaxel into regions where the dominant ionizing source is star formation, composite, or AGN using the BPT \citep{bal81} diagnostic based on the [OIII]/\hb~vs. [NII]/\ha~ line ratios \citep{kew01,kew06}. 
We then identify star-forming spaxels to be those satisfying both BPT-classified star-forming criteria and the \ha~equivalent width (EW) $>$ 6 \AA~cut \citep[e.g.,][]{cid11}. The retired spaxels are selected to be those having \ha~ EW $<$ 3 \AA~ and S/N $>$ 3 in \ha~ \citep[see also][]{ell21b}. The remaining spaxels are categorized as "other" spaxels. These could be either star-forming spaxels but with very low \ha~EW, composite regions, or non-star-forming regions. As we will see later, the behaviors of these `other' spaxels are similar to the retired spaxels, indicating that they are predominantly regions with suppressed star formation.

\begin{figure*}
\centering
\includegraphics[angle=0,width=0.95\textwidth]{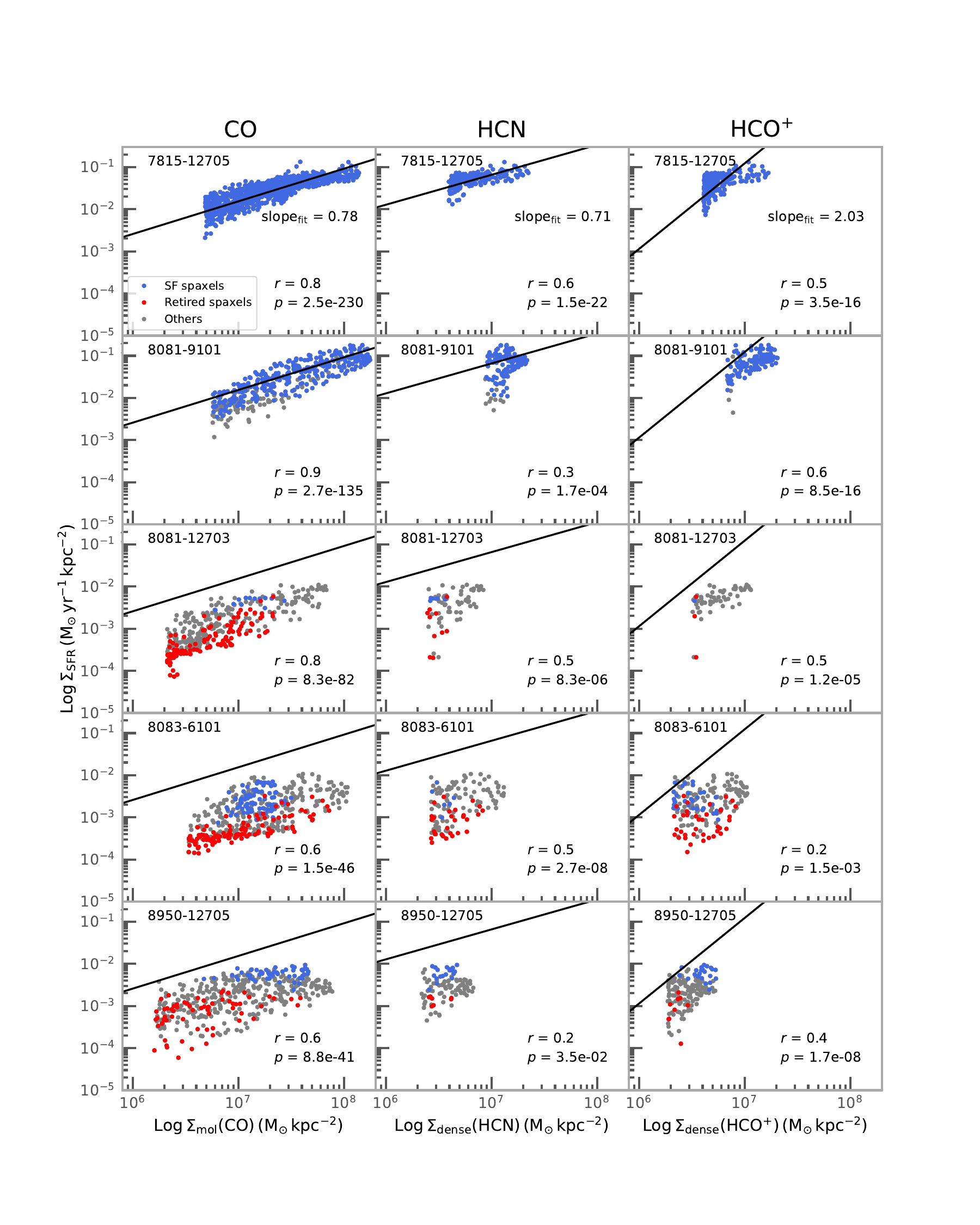}
\caption{The resolved Schmidt-Kennicutt relation (i.e., SFR surface density vs. gas mass surface density) using various gas tracers. From left to right: CO, HCN, and \HCO.  The top two rows are for main sequence galaxies and the bottom 3 rows are for green valley galaxies. Each data point represents the measurement from individual spaxels. The sharp edges of the data points in the low end of gas mass surface density correspond to the 2.5-$\sigma$ detection limits. The star-forming spaxels, retired spaxels, and the remaining spaxels are shown by blue, red, and grey symbols, respectively. At a given gas tracer, the black line represents the best fit of the star-forming spaxels of the MS galaxy 7815-12705 and is the same for all other panels corresponding to a particular gas tracer to guide the eyes. The values of the Pearson ($r$) correlation coefficients and $p$-values computed using all spaxels are reported in each panel.
\label{fig:sfr-M}}
\end{figure*}

\section{RESULTS}
\subsection{Dense gas Schmidt-Kennicutt Relation\label{sec:sk_dense}}
We begin by showing the resolved molecular gas Schmidt-Kennicutt relation, i.e., the star formation rate surface density vs. the molecular gas surface density, for the 2 MS (upper 2 rows) and 3 GV (lower 3 rows) galaxies separately in Figure \ref{fig:sfr-M}. The gas masses converted based on CO, HCN, and \HCO~ luminosities are shown from the left to the right panels. Each dot on the plot represents measurements for an individual spaxel. The blue symbols represent spaxels classified as star-forming, the red points denote spaxels classified as retired, and the gray points come from the remaining spaxels. It can be seen that (as expected) the two MS galaxies are dominated by the star-forming spaxels, whereas the 3 GV galaxies primarily consist of retired and `other' spaxels.

The Pearson correlation coefficients ($r$) are computed using all types of spaxels and are displayed in each panel. For the main sequence galaxy 7815-12705, \sigsfr~ is found to correlate with all the three gas tracers (CO, HCN, and \HCO), although the correlation for \HCO~ is only moderate ($r$ = 0.5). Similar behavior is also seen for the other main sequence galaxy 8081-9101. However, in this case the correlation for HCN is weaker compared to either CO or \HCO, likely affected by the narrow dynamical range in the \sigdense. The correlation coefficients in the CO-based SK relation are in general greater for MS than for GV galaxies. This trend is however less obvious for HCN or \HCO-based SK relations.

To guide our eyes when comparing objects-to-objects and spaxel types-to-spaxel types, we fit the star-forming spaxels of the main sequence galaxy 7815-12705 with the orthogonal distance regression (ODR) fitting method for the three gas tracers separately. The best fits of the object 7815-12705 are shown in black as a reference line in the panels of the other objects as well.  The distribution for the retired and other spaxels shows a large scatter and lies toward a lower \sigsfr~at a given gas surface density when compared to star-forming spaxels regardless of the gas tracer. 
It is also apparent that even for the star-forming spaxels, \sigsfr~ in GV galaxies are systematically lower than those in MS galaxies at a given gas surface density. This is in line with the finding that GV galaxies deviate from the scaling relations formed by MS galaxies not only for retired spaxels but also for star-forming spaxels \citep{lin19b,bro20,lin22}. 

\begin{figure*}
\centering
\includegraphics[angle=0,width=0.95\textwidth]{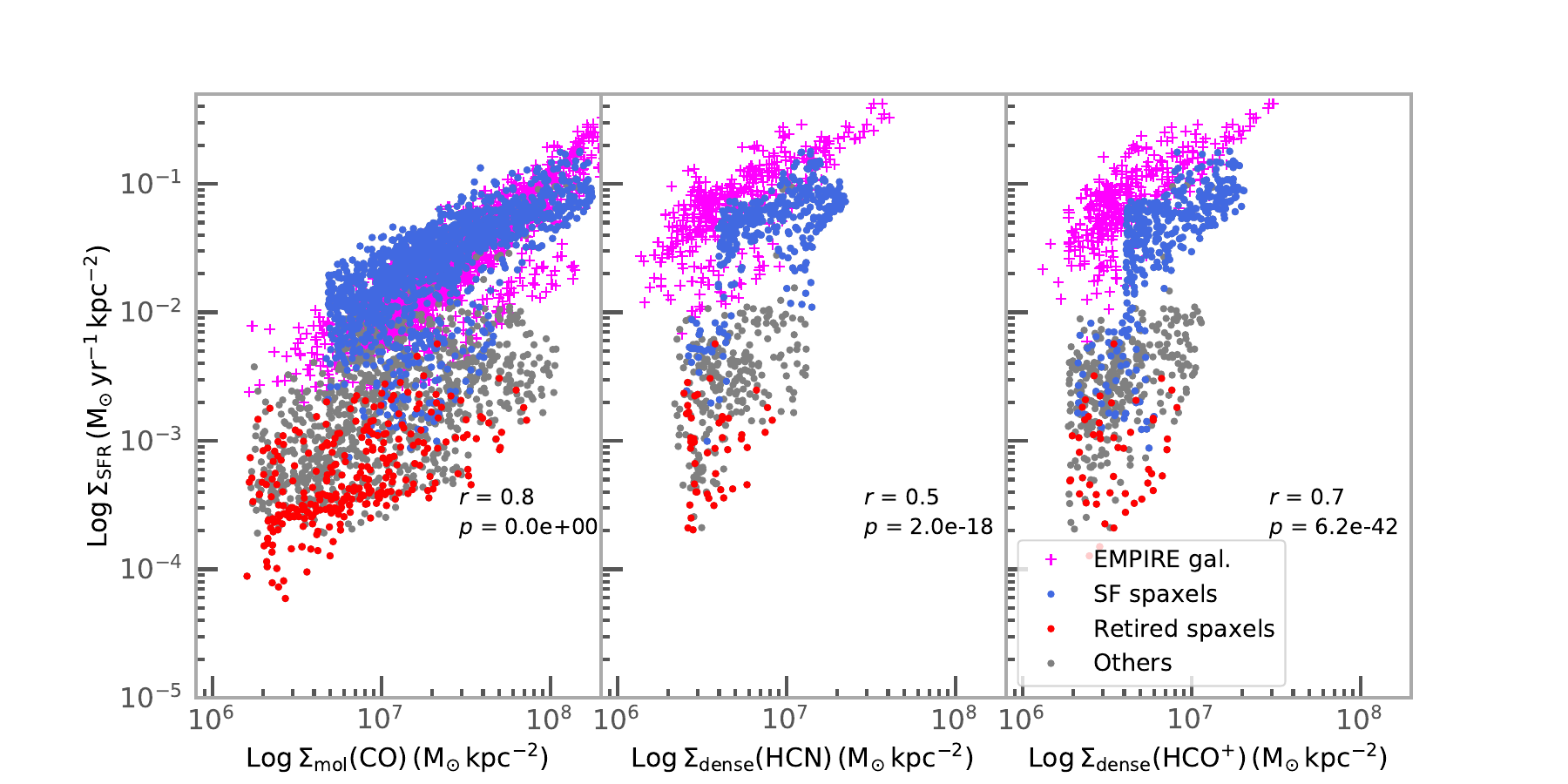}
\includegraphics[angle=0,width=0.95\textwidth]{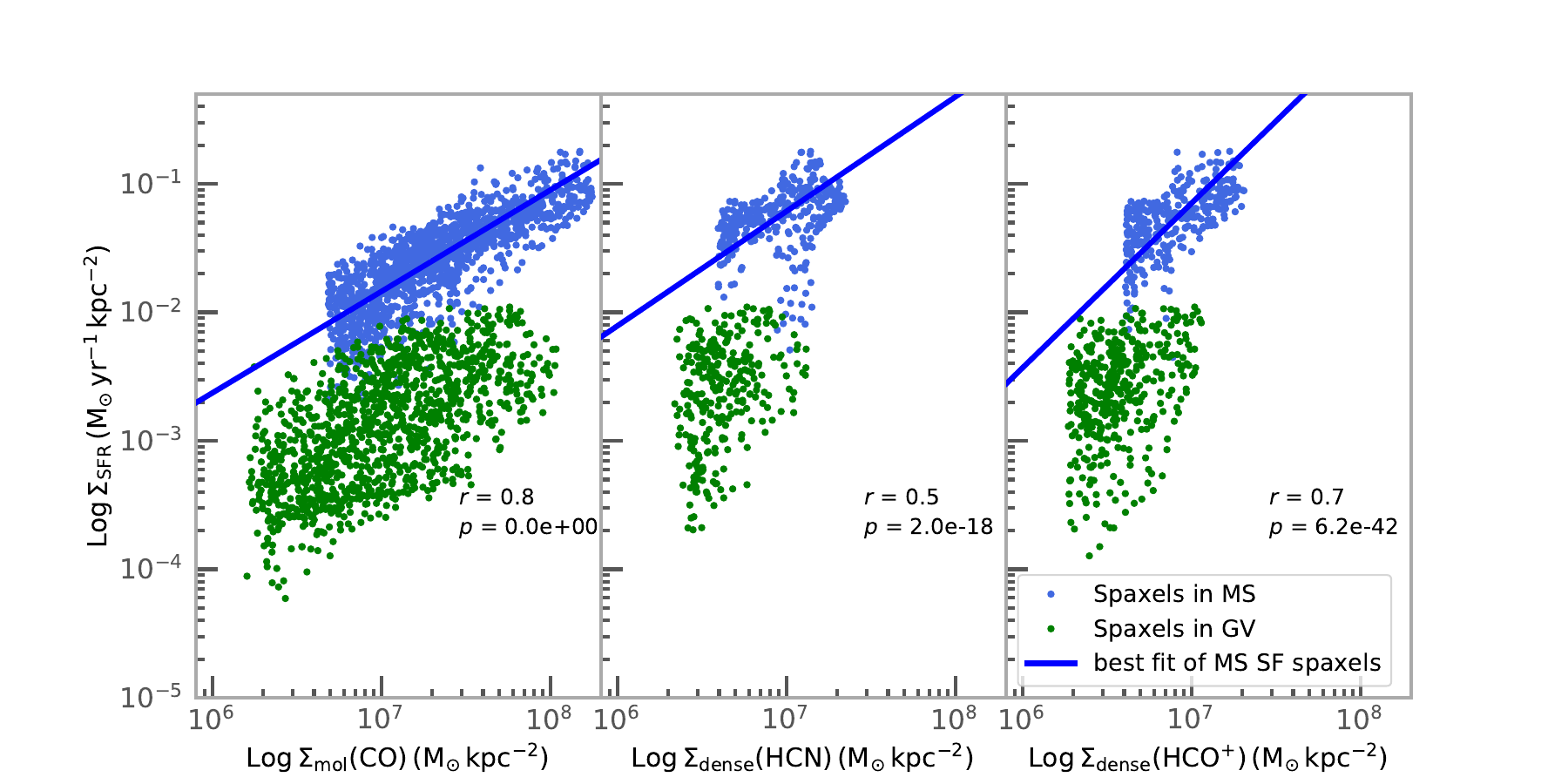}

\caption{
The resolved Schmidt-Kennicutt relation (i.e., SFR surface density vs. gas mass surface density) using various gas tracers, with all galaxies plotted together. The sharp edges of the data points in the low end of gas mass surface density correspond to the 2.5-$\sigma$ detection limits.  From left to right: CO, HCN, and \HCO.  The magenta line represents the best fit from the EMPIRE sample \citep{jim19} for comparison. Upper panels: The star-forming spaxels, retired spaxels, and the remaining spaxels are shown by blue, red, and grey symbols,
respectively.  Bottom panels: The spaxels located in the main sequence (MS) and green valley galaxies (GV) are shown by blue and green symbols, respectively. The scatters of the star-forming spaxels relative to the fit are labeled in the lower right corner of each panel. In all panels, the Pearson correlation coefficients ($r$) and $p$-values are computed using star-forming spaxels belonging to MS galaxies only.
\label{fig:sfr-M_all}}
\end{figure*}

Figure \ref{fig:sfr-M_all} is similar to Figure \ref{fig:sfr-M} but now we show the spaxel distributions by combining all the 5 galaxies together. In the upper panel, the data points are color-coded by their spaxel types (blue: star-forming; red: retired; others: grey). For comparison, we also plot the EMPIRE sample (magenta points), which consists of 9 galaxies with resolved measurements ($\sim$1-2 kpc scale resolution) from \citet{jim19}. We see that the CO-based SK relation for ALMaQUEST star-forming spaxels is in good agreement with the EMPIRE data points (the upper left panel). On the other hand, the ALMaQUEST HCN (the upper middle panel) and \HCO~(the upper left panel) SK relation of the star-forming spaxels are offset from those of the EMPIRE sample toward a lower value of \sigsfr~at a given gas mass surface density. It is worth mentioning that the SFR from the EMPIRE sample is based on the total infrared luminosity whereas our sample is converted from the extinction-corrected \ha~luminosity. However, the difference in the star formation rate tracers does not seem to be responsible for the systematic difference seen in the dense gas SK relation between our samples since there is a good agreement in the CO-based SK relation. We note that there are in fact significant galaxy-to-galaxy variations even within the EMPIRE sample itself \citep{jim19} and the ALMaQUEST star-forming spaxels overlap with some of the EMPIRE data points. This suggests that the difference can be intrinsic rather than systematic in the \sigsfr~measurements. There are several other speculations that might explain the offset in dense gas SK relation and hence \SFEdense~ between EMPIRE star-forming galaxies and the MS targets in this study. These include the potential variations of $\alpha_{\rm HCN}$ and/or $\alpha_{\rm HCO^{+}}$ (in which case the dense gas conversion factors are overestimated when adopting fixed values), intrinsic differences in \SFEdense~ within the star-forming main sequence galaxies, and calibration issues. Given that much less is known about the dense gas conversion factors in any kind of extragalactic system, we cannot be certain that the fixed conversion factors hold for the galaxies in this study or even for EMPIRE. Nevertheless, in the case the SK relation of our MS sample is not representative, unlike EMPIRE galaxies, the offset in \SFEdense~ at a given \sigdense~ between typical star-forming galaxies (as inferred by EMPIRE) and the green valley galaxies is even larger than what we conclude in this work. 

Within the ALMaQUEST sample, it is clear that the star-forming and retired spaxels occupy distinct parameter spaces with each other. The non-star-forming (retired plus other) regions do not follow the same scaling relation as the star-forminng spaxels and possess lower SFE. While this feature is already known when adopting the CO tracer from earlier works \citep[e.g.,][]{lin22}, the fact that the same trend also exhibits for the dense gas tracers is not fully expected. The deviation from the dense gas Schmidt-Kennucutt relation basically suggests that the level of star formation activities is not entirely regulated by the amount of available dense gas. Other physical conditions also play roles in determining whether the dense gas is able to form stars or not. Several earlier studies \citep{gao04,wu05,lad10} proposed a linear correlation between HCN and SFR, indicating a fixed value of \SFEdense, spanning scales from clouds to galaxies. However, more recent investigations have revealed significant variations in the \SFEdense, across diverse galactic environments and cloud properties \citep[e.g.,][]{lon13,use15,jim19,que19,san22,neu23}. Our findings align with this notion, demonstrating that \SFEdense~is not a constant and varies with the spaxel type.

In the bottom panels of Figure \ref{fig:sfr-M_all}, we color code the spaxels by the types of their host galaxies (blue: MS; green: GV). As expected, MG and GV occupy distinct parameter spaces in this diagram since MS galaxies are dominated by star-forming spaxels whilst GV galaxies comprise primarily the non-star-forming spaxels. Therefore, we conclude that the 3 GV galaxies have distinct \SFEmol~and \SFEdense~from the 2 MS galaxies despite that they have similar \fmol~by selection.

\subsection{Is the molecular gas SFE dependent on the dense gas fraction?\label{sec:sfe-fgas}}
Equation \ref{eq:sfe} suggests that the overall molecular gas star formation efficiency (\SFEmol) could depend on two things, the dense-to-molecular gas ratio (\fdense) and the dense gas star formation efficiency (\SFEdense). We now first investigate whether \SFEmol~ is primarily driven by \fdense~ or not. If the answer is `yes', we should expect a positive correlation
between \SFEmol~ and \fdense~.

\begin{figure}
\centering
\includegraphics[angle=0,width=0.45\textwidth]{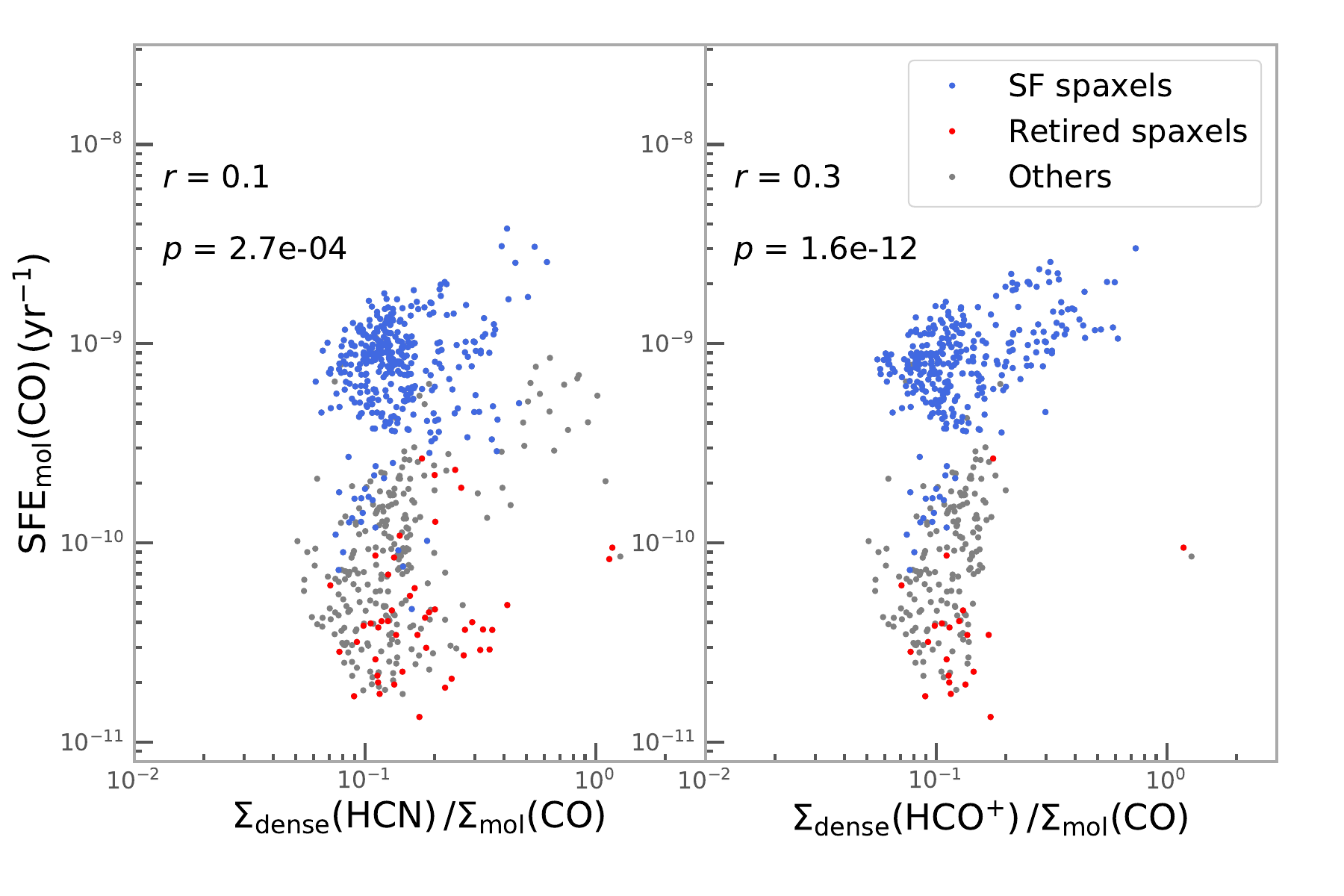}
\includegraphics[angle=0,width=0.45\textwidth]{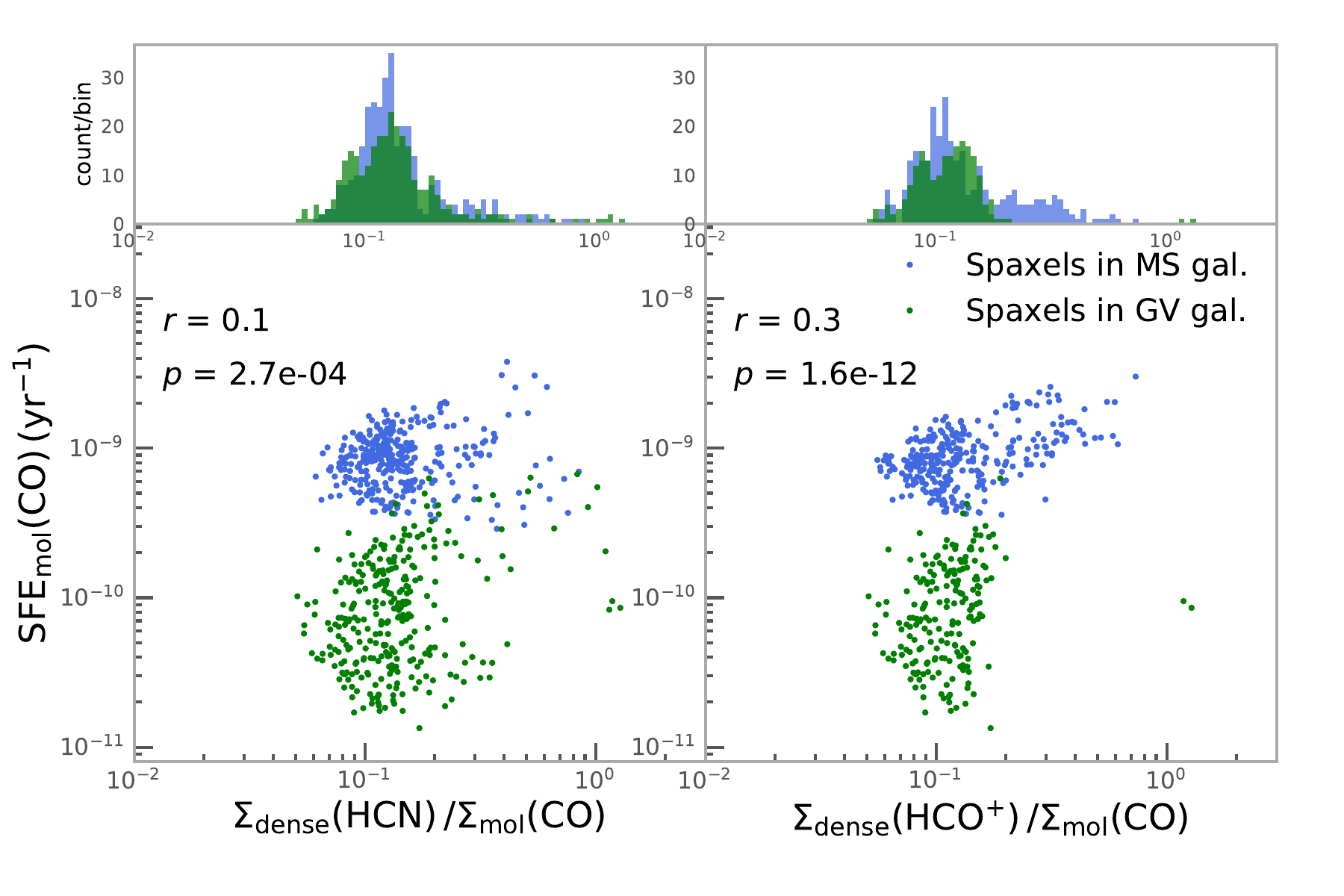}

\caption{
The molecular gas-based SFE$_{mol}$ against the dense-to-molecular mass ratio using HCN (left panel) and the dense-to-molecular mass ratio using \HCO~(right panel). The Pearson correlation coefficients ($r$) and $p$-values computed using all spaxels are reported in each panel. Top panels: the star-forming, retired, and remaining spaxels are shown in blue, red, and grey colors, respectively. Bottom panels: spaxels in MS and GV galaxies are shown in blue and green colors, respectively. Histograms of \fdense~ for both MS and GV galaxies are presented in the upper x-axes. 
\label{fig:sfe-fgas}}
\end{figure}

The upper panels of Figure \ref{fig:sfe-fgas} shows \SFEmol~against \fdense, computed using the HCN (left panel)
and \HCO~ (right panel). There is a weak correlation between the x- and y- axes for the star-forming spaxels, indicating that \SFEmol~ is mildly connected to the dense gas fraction in star-forming regions. Turning next to the non-star-forming regions, we see that these spaxels have lower \SFEmol~ than the star-forming spaxels as expected based on the work of \citet{lin22}. However, it is notable that, despite their very different \SFEmol, the two types of spaxels (star-forming vs. non-star-forming) share a similar range of \fdense.
This implies that the low \SFEmol~ that leads to quenching in non-star forming regions is not due to the lack of dense gas.

In the bottom panels of Figure \ref{fig:sfe-fgas}, spaxels are color-coded based on the types of galaxies they belong to: the main sequence (MS; blue points) and green valley (GV; green points) galaxies. We also plot the histograms of \fdense~for the two types of galaxies. As mentioned earlier, the majority of spaxels in MS galaxies correspond to star-forming regions, while those in GV galaxies are predominantly non-star-forming spaxels. As a result, it is evident from the plot that while the spaxels in MS and GV galaxies exhibit distinct distributions in terms of \SFEmol, they demonstrate similar distributions of \fdense, emphasizing that \fdense~ does not play a significant role in determining the difference in \SFEmol~between MS and GV galaxies. This phenomenon is similar to what have been found in some gas-rich early-type galaxies \citep{cro12} and a certain fraction of post-starburst galaxies \citep{fre23}. We will return to this issue in \S\ref{sec:literature}.

\subsection{Is the molecular gas SFE dependent on the dense gas SFE?\label{sec:sfe-sfe}}
In the previous sub-section, we found that there was only a weak correlation between \SFEmol~ and \fdense.  We, therefore, suggested that a similar range of dense gas fractions between star-forming and non-star-forming spaxels supports differences in dense gas star formation efficiency as the primary quenching mechanism.  In this sub-section, we test this hypothesis directly.

\begin{figure}
\centering
\includegraphics[angle=0,width=0.45\textwidth]{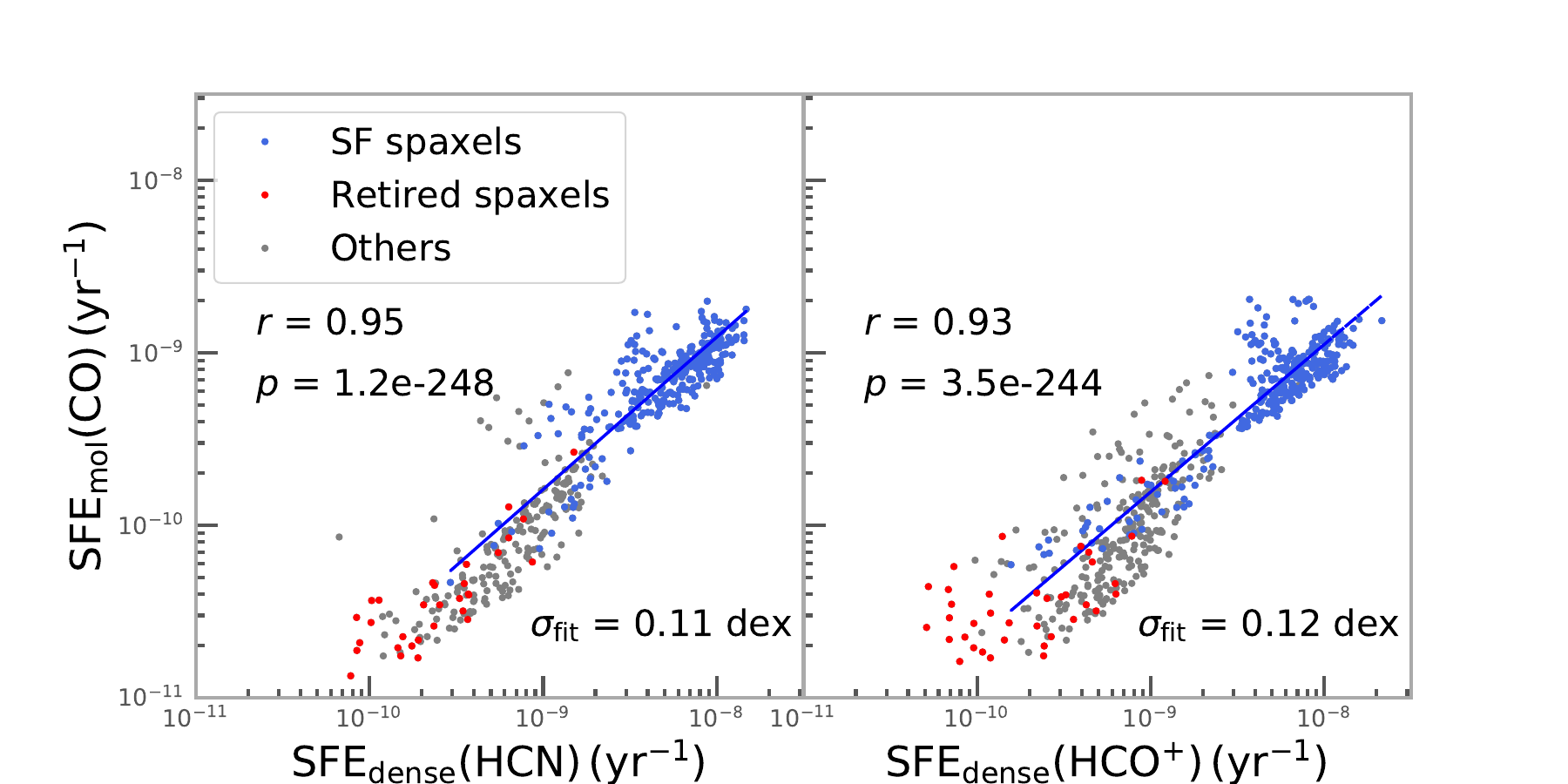}
\includegraphics[angle=0,width=0.45\textwidth]{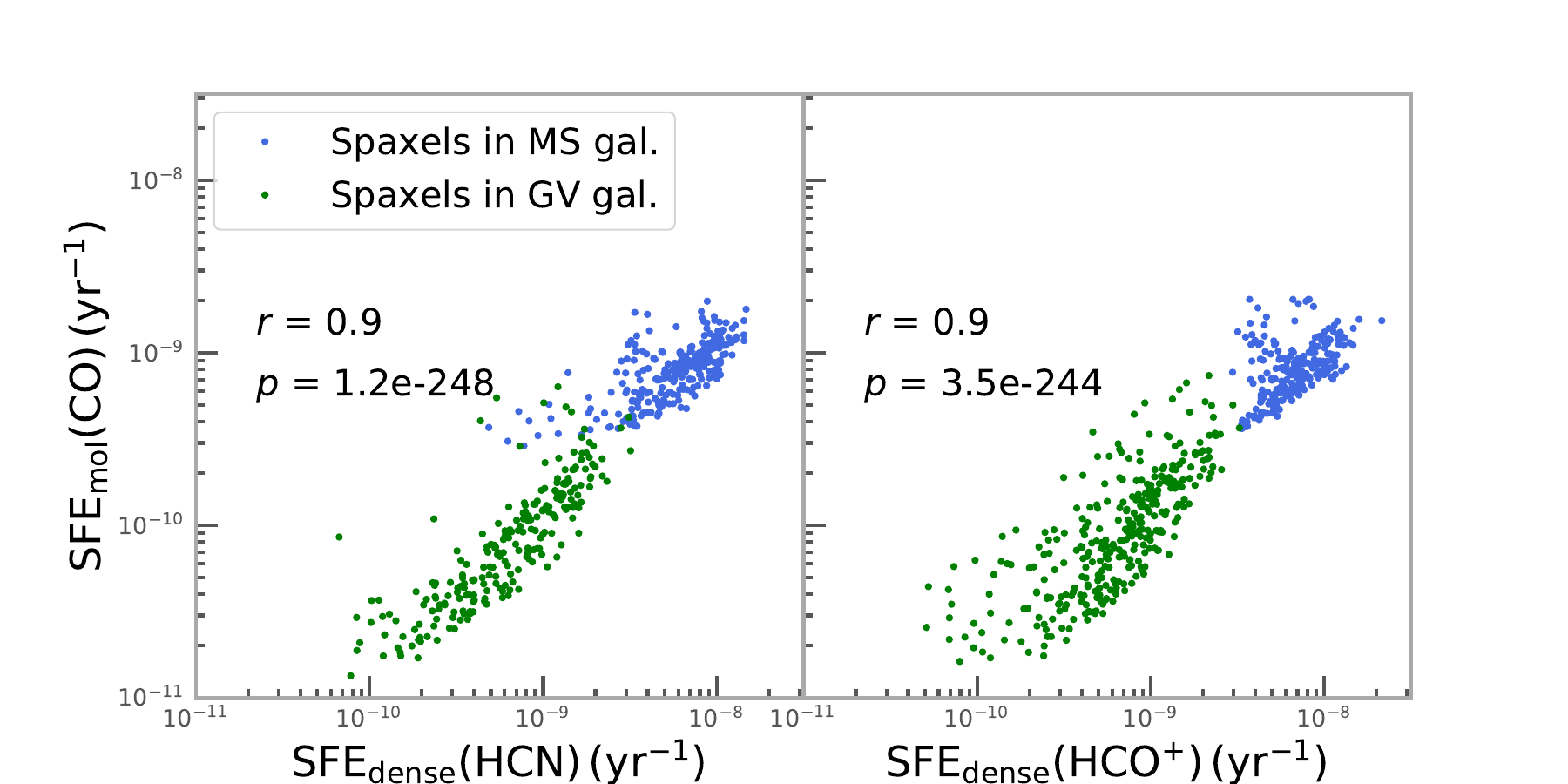}

\caption{
The molecular gas-based star formation efficiency (\SFEmol) against the dense gas-based star formation efficiency (\SFEdense) that are measured using HCN (left panel) and \HCO~(right panel). 
The Pearson correlation coefficients ($r$) and $p$-values computed using all spaxels are reported in each panel. Top panels: The star-forming, retired, and remaining spaxels are shown in blue, red, and grey colors, respectively. The blue line represents the best fit of the star-forming spaxels from all 5 objects. The scatters of the star-forming spaxels relative to the fit are labeled in the lower right corner of each panel. Bottom panels: spaxels in the main sequence and green valley galaxies are color-coded by blue and green, respectively. \label{fig:sfe-sfe}}
\end{figure}

In the upper two panels of Figure \ref{fig:sfe-sfe}, we show the molecular gas-based SFE as a function of the dense gas-based SFE (left: HCN; right: \HCO). It is obvious that there is a strong dependence of \SFEmol~on \SFEdense, as indicated by the high correlation coefficients (0.93-0.95) even when all types of spaxels are considered together regardless of whether they are star-forming spaxels or not. The ODR best fits for star-forming spaxels are shown in blue lines as a reference. Interestingly, it can be seen that the non-star-forming spaxels also follow a similar trend to the one that is formed by the star-forming spaxels, unlike the bimodal distribution on the \SFEmol~and \fdense~ plane between the star-forming and non-star-forming spaxels as seen in Figure \ref{fig:sfe-fgas}. 

In the bottom panels of Figure \ref{fig:sfe-sfe}, we present similar plots but this time we color code the spaxels according to their host galaxies: blue for main sequence galaxies and green for green valley galaxies. It can be seen that the GV spaxels extend the distribution of MS spaxels towards lower values of \SFEmol~ and \SFEdense. This again reflects the fact that GVs in our sample are dominated by non-star-forming spaxels in contrast to main sequence galaxies, which consist primarily star-forming spaxels. 

One caveat of the correlation analysis when discussing the relative importance between \fdense~and \SFEdense~in determining \SFEmol~is that \SFEmol~and \fdense~share the same denominator (i.e., \sigmol) while \SFEdense~and \SFEmol~share the same nominator (i.e., \sigsfr), which can lead to artificial correlations in both cases. To test this effect, we generate synthesized datasets by shuffling \sigmol~and \sigdense~and repeat the correlation analysis in \S \ref{sec:sfe-fgas} on the \SFEmol~vs. \fdense~ relation. The procedure is then repeated 500 times. The median Pearson correlation coefficient $r$ from 500 trials is increased to $r = 0.36$ and $r = 0.34$ using the synthesized data for HCN and \HCO, respectively, which is slightly greater than the correlation of the real datasets ($r = 0.1-0.3$). We can therefore robustly conclude that there is no apparent dependence on \fdense. 

Similarly, we test this effect on the \SFEmol~and \SFEdense~relation by shuffling both \sigmol~and \sigsfr. This time, the median correlation coefficients from 500 trials are already found to be high, $\sim$ 0.87 and 0.85 for the case of HCN and \HCO, respectively, only slightly lower than the correlations when using the real physical dataset ($r$ = 0.93 for HCN and 0.95 for \HCO), making it difficult to interpret the results. As the artificial correlation arises due to the stretch of the common nominator (\sigsfr~in this case), they are sensitive to the dynamical range in \sigsfr. We therefore conduct another experiment, by only selecting data points in the central half of the \sigsfr~range, specifically -3 < Log$_{10}$\sigsfr~< -1.5, to mitigate this effect. In this case, the correlations for the real datasets are 0.85 (HCN) and 0.84 (\HCO) and those for the synthesized datasets are 0.70 (HCN) and 0.65 (\HCO). The difference becomes more apparent in this subset of sample. While it remains difficult to quantify the intrinsic correlations, the central point is that the correlation  between \SFEmol~and \SFEdense~ is stronger when using the real physical dataset than in the synthesized data, in contrast to the worse correlation between \SFEmol~vs. \fdense~ in the real datasets than that in the synthesized data. This hints that the dependence of \SFEmol~on \SFEdense~has more physical connection than on \fdense~in a relative sense. Therefore, the low \SFEdense~ is more likely the primary driver of the low \SFEmol~ in GV galaxies.

\begin{figure*}
\centering
\includegraphics[angle=0,width=0.95\textwidth]{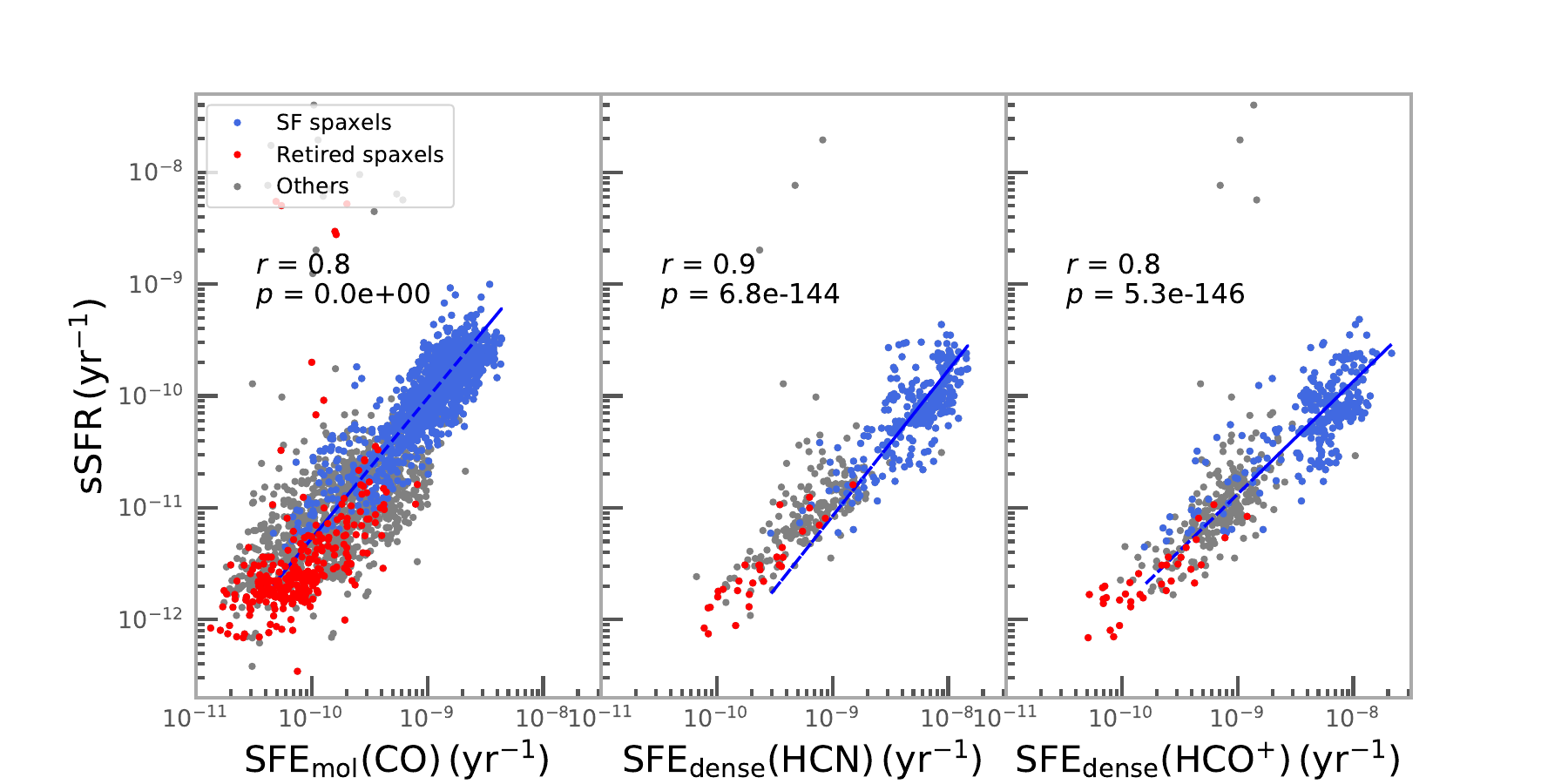}

\caption{The specific star formation rate against the molecular gas-based SFE measured using CO (left panel), dense gas-based SFE measured using HCN (middle panel), and dense gas-based SFE measured using \HCO~(right panel). The star-forming, retired, and remaining spaxels are shown in blue, red, and grey colors, respectively. The blue line represents the best fit of the star-forming spaxels from all objects. The Pearson correlation coefficients ($r$) and $p$-values computed using all spaxels are reported in each panel.  \label{fig:ssfr-sfe}}
\end{figure*}

\begin{figure*}
\centering
\includegraphics[angle=0,width=0.95\textwidth]{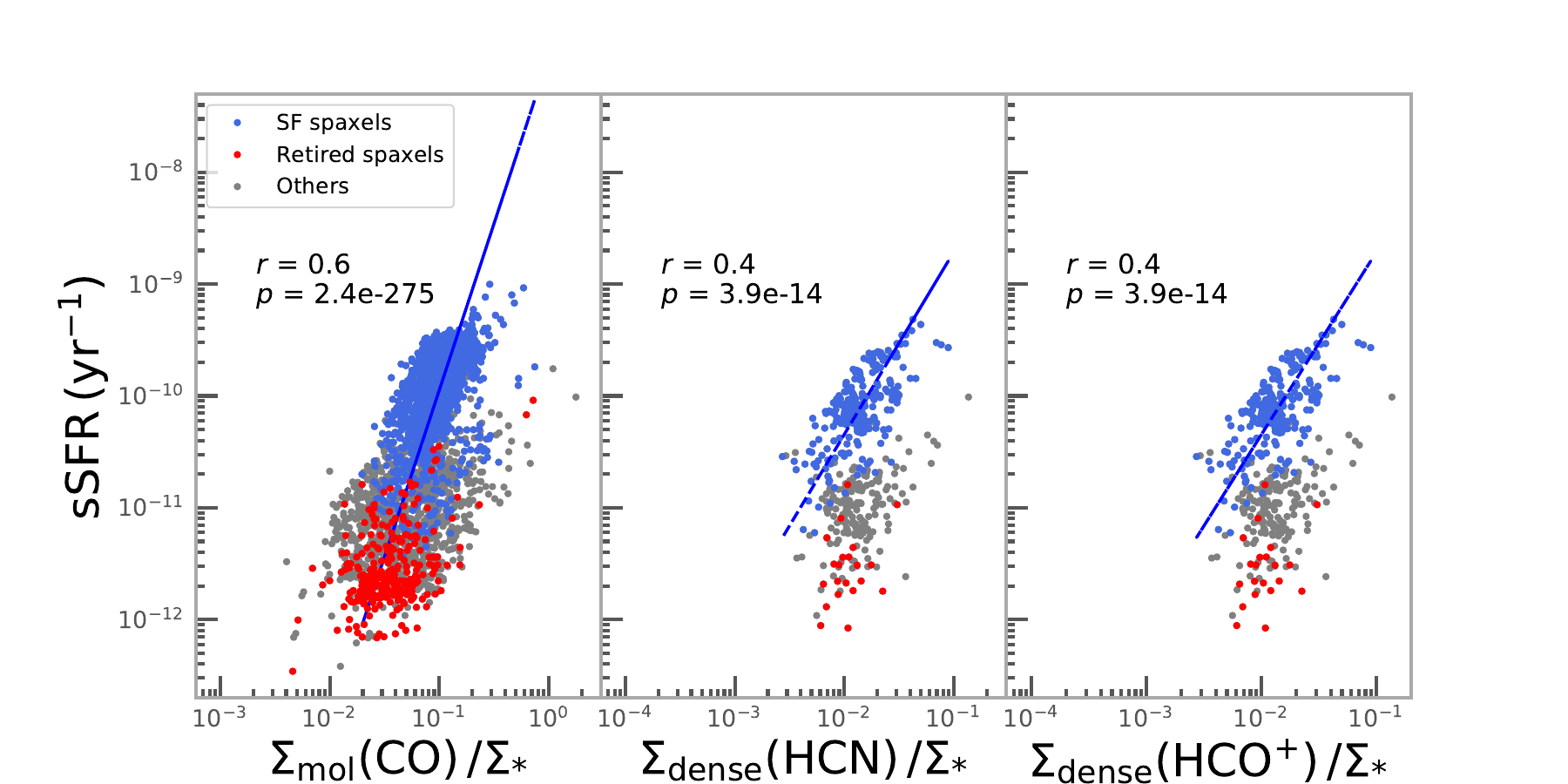}

\caption{The specific star formation rate (sSFR) against the molecular gas-to-stellar mass ratio (left panel), dense gas-to-stellar mass ratio using HCN (middle panel), and dense gas-to-stellar mass ratio using \HCO~(right panel). The star-forming, retired, and remaining spaxels are shown in blue, red, and grey colors, respectively. The blue line represents the best fit of the star-forming spaxels from all objects. The Pearson correlation coefficients ($r$) and $p$-values computed using all spaxels are reported in each panel.  \label{fig:ssfr-gastosm}}
\end{figure*}

\begin{figure*}
\centering
\includegraphics[angle=0,width=0.95\textwidth]{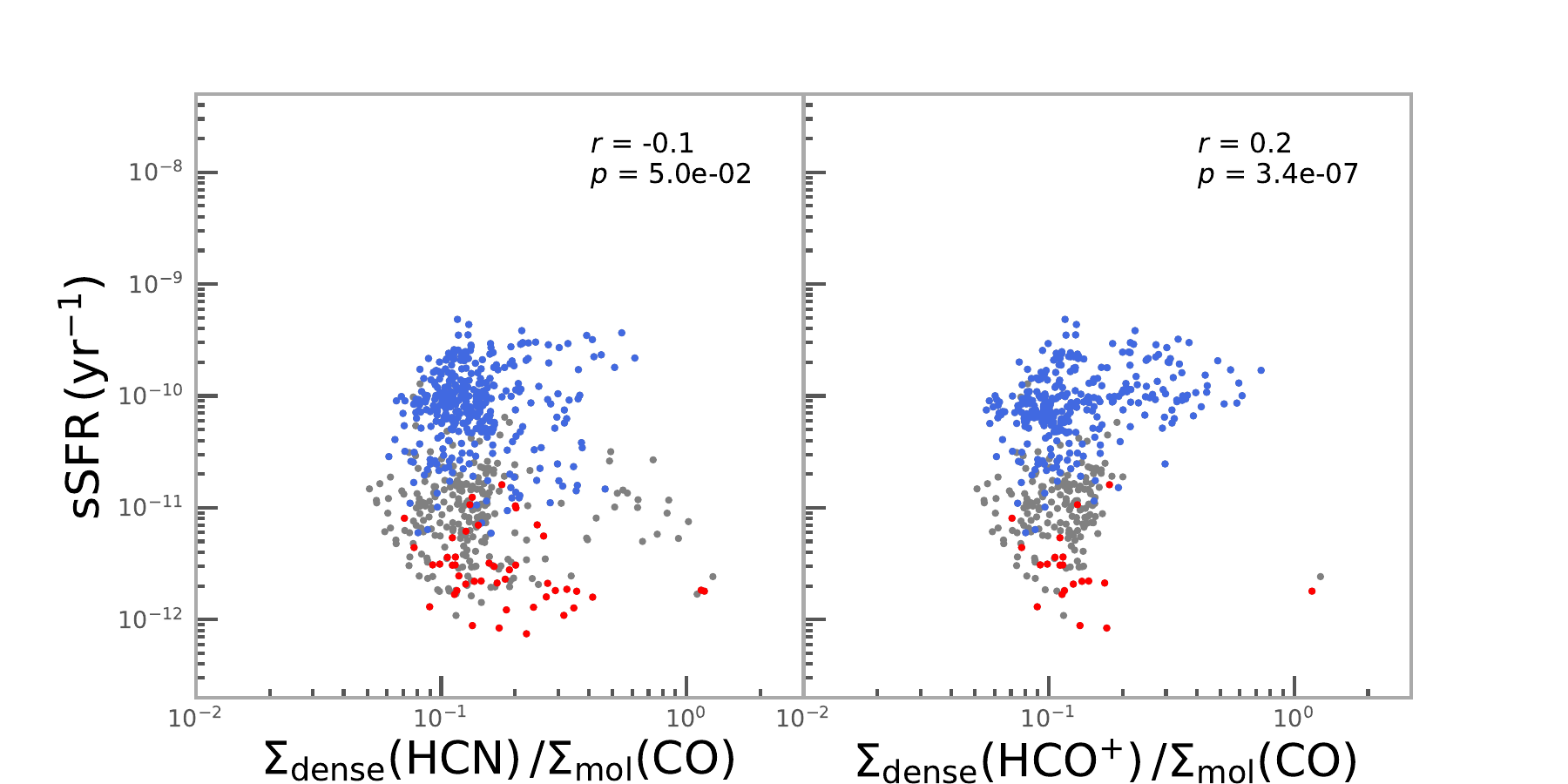}

\caption{The specific star formation rate (sSFR) against the dense-to-molecular gas mass ratio measured using HCN (left panel) and \HCO~(right panel). The star-forming, retired, and remaining spaxels are shown in blue, red, and grey colors, respectively. The Pearson correlation coefficients ($r$) and $p$-values computed using all spaxels are reported in each panel.  \label{fig:ssfr-gas}}
\end{figure*}

\subsection{sSFR vs SFE with different tracers}
Having discussed the relative importance between \fgas~and \SFEdense~in determining \SFEmol, we now turn to one of the central questions of this work, which is to identify the major mechanism that is responsible for lowering the specific star formation rate in galaxies. Recalling that sSFR is the product of SFE and \fgas, i.e., sSFR $=$ \fgas~$\times$SFE, in which \fgas~and SFE could be either expressed using the overall molecular gas or dense molecular gas. It is possible to differentiate the contributions between the gas abundance and star formation efficiency by investigating the dependence of sSFR on \fgas~and SFE. Here we express the sSFR as the following:

\begin{eqnarray}\label{eq:ssfr1}
\rm sSFR &=& \Sigma_{\rm SFR}/\Sigma_{*}\nonumber\\
&=& \Sigma_{\rm SFR}/\Sigma_{\rm dense} \times \Sigma_{\rm dense}/\Sigma_{*}\nonumber\\
&=& \rm SFE_{\rm dense} \times R_{\rm dense},
\end{eqnarray}
where \Rdense~is the ratio of the dense gas mass to the stellar mass. This should not be confused with the dense gas fraction (\fdense) defined previously, which refers to the dense gas to the molecular gas mass ratio.

We first look at the dependence of sSFR on SFE using various tracers, including CO(1-0), HCN, and \HCO. Again in the well-accepted scenario that the stars are formed within the dense molecular clouds, we shall expect a tighter link between sSFR and HCN- or \HCO- based SFE than that between sSFR and CO-based SFE. Figure \ref{fig:ssfr-sfe} plots the relations of sSFR against the SFE defined with various tracers (CO, HCN, and \HCO). We see that in every case there is a strong correlation between sSFR and SFE, regardless of the tracers being used. The correlation coefficients are comparable between the CO-based SFE and the HCN- (or \HCO-based) SFE. Another interesting feature is that the non-star-forming spaxels also follow the same trend as the one formed by the star-forming spaxels, suggesting that SFE plays a key role in determining sSFR across the whole range of star formation levels from active to quiescent phases.

On the other hand, the correlation between sSFR and the gas-to-stellar mass ratio is much weaker (see Figure \ref{fig:ssfr-gastosm}) compared to the relation between sSFR and SFE reported in Figure \ref{fig:ssfr-sfe}. When we consider all types of spaxels together (including star-forming and non-star-forming spaxels), the correlation is only apparent in the case of CO, in which the non-star-forming spaxels also lie on the trend formed by the star-forming spaxels. On the other hand, in the case of HCN or \HCO, the non-star-forming spaxels show a departure from the sSFR vs. gas-to-stellar mass ratio that is formed by the star-forming spaxels. In addition, the non-star-forming spaxels share a similar range in the dense gas-to-stellar mass ratio as the star-forming spaxels. Our results suggest that while the dense gas-to-stellar mass ratio does play a role in regulating the sSFR for star-forming spaxels, it is not a dominant factor in determining the sSFR in regions that are undergoing quenching processes.

Equation \ref{eq:ssfr1} can also be further expanded as the following:

\begin{eqnarray}\label{eq:ssfr2}
sSFR &=& SFR/M_{dense} \times M_{dense}/M_{mol} \times M_{mol}/M_{*}\nonumber\\
&=&   SFE_{dense} \times f_{dense} \times f_{mol}.
\end{eqnarray}
Therefore, we also investigate the relation between sSFR and \fdense, which is shown in Figure \ref{fig:ssfr-gas}. It can be seen that there is no apparent dependence of sSFR against \fdense~ in either the case of HCN or \HCO. In summary, among several parameters we have explored, the dependence of sSFR on \SFEdense~ (and/or \SFEmol) is stronger than on \fdense~or \fmol, suggesting the star formation efficiency is the dominant factor determining the kpc-scale sSFR.

\begin{figure*}
\centering
\includegraphics[angle=0,width=0.95\textwidth]{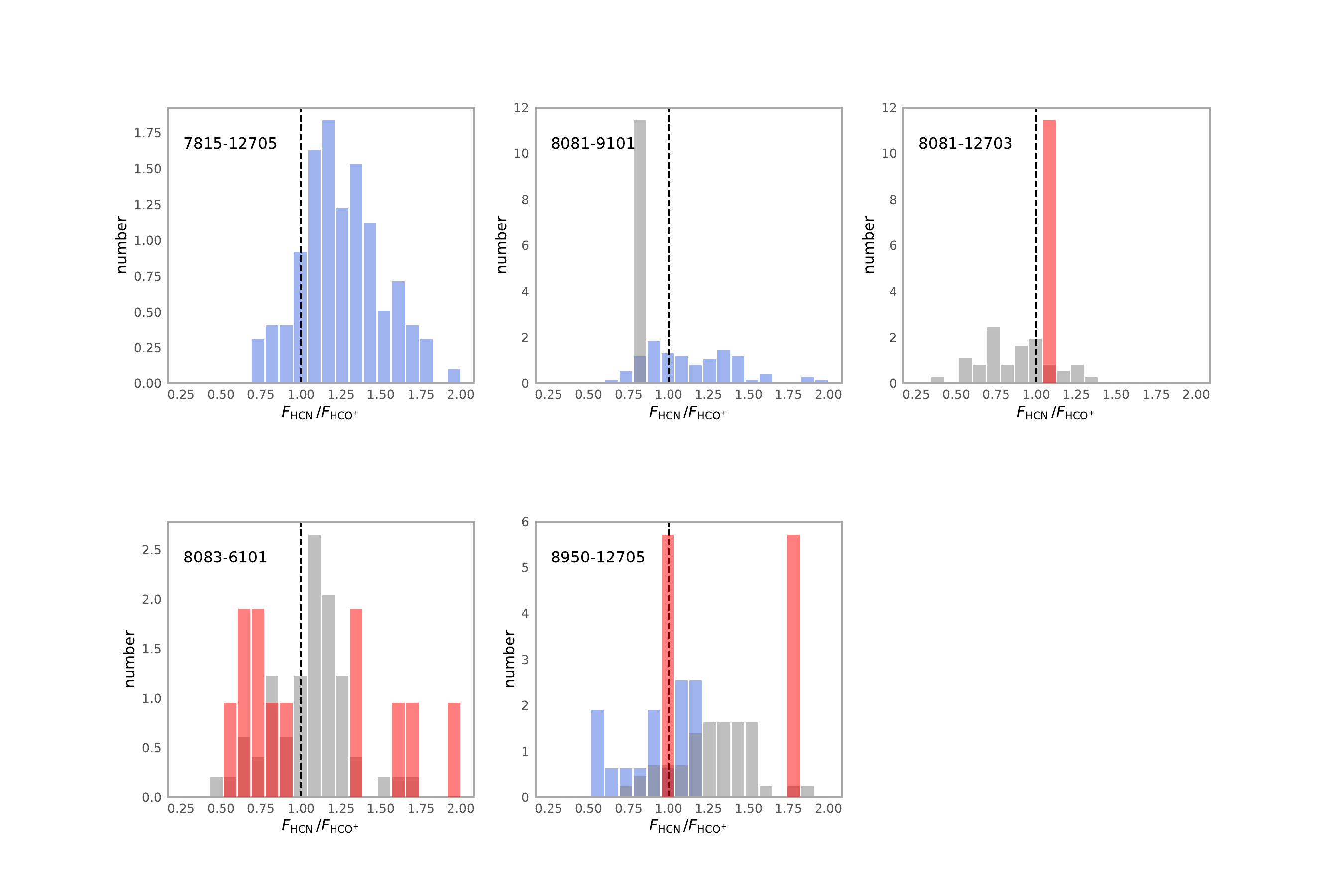}

\caption{
The HCN-to-\HCO~flux ratio of the 5 galaxies analyzed in this work. The star-forming, retired, and the remaining spaxels are shown in blue, red, and grey colors, respectively. The vertical dashed line represents the HCN-to-\HCO~flux ratio = 1 to guide the eyes.
\label{fig:his_fhcnhco}}
\end{figure*}

\begin{figure*}
\centering
\includegraphics[angle=0,width=0.95\textwidth]{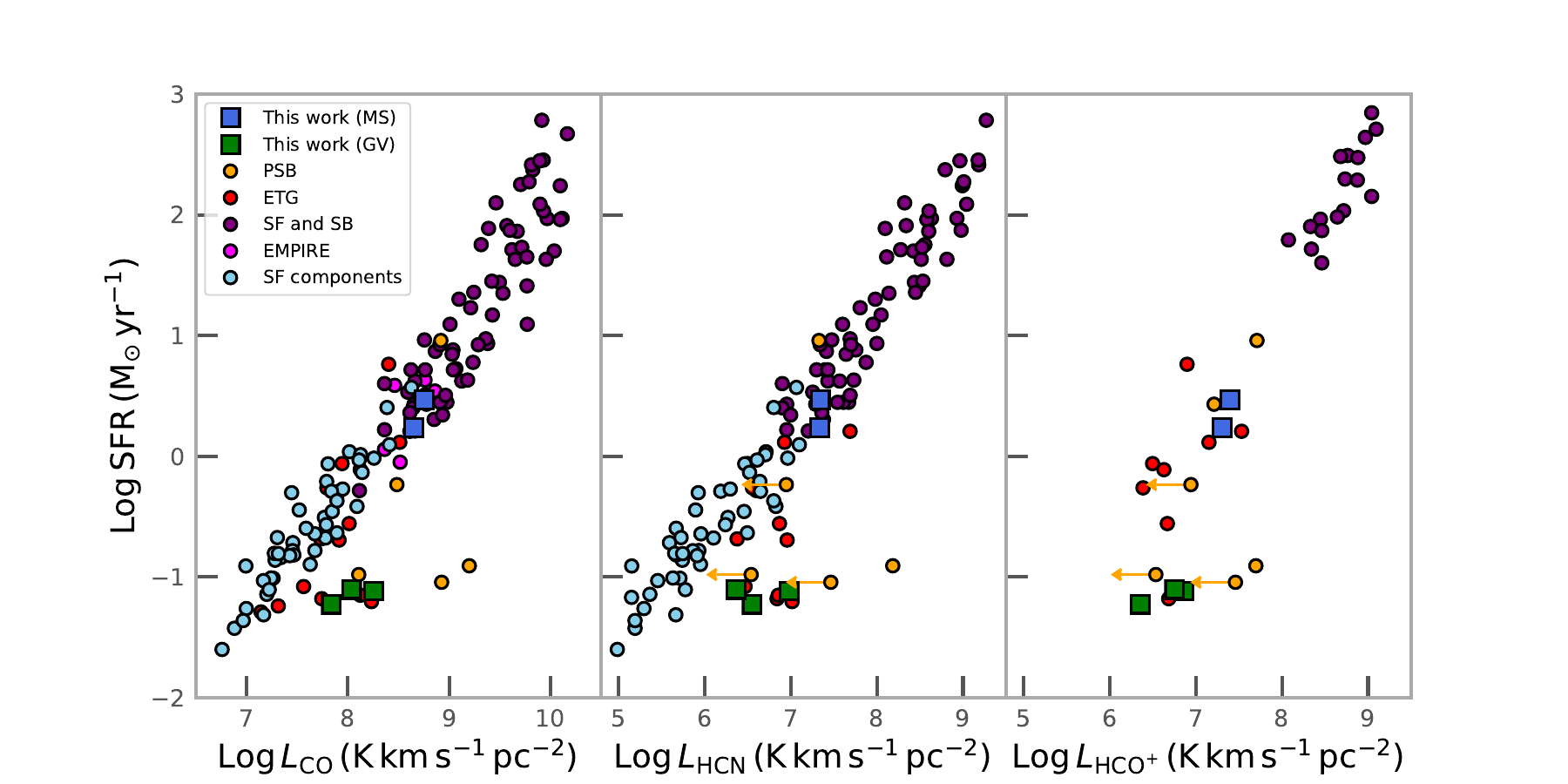}

\caption{The (global) SFR versus $L_{gas}$ relation for ALMaQUEST galaxies used in this work (main-sequence: blue squares; green valley: green squares), post-starburst galaxies \citep{fre23}, EMPIRE star-forming galaxies \citep{jim19}, ETGs \citep{cro12}, star-forming and starburst galaxies \citep{gao04}, and subcomponents of star-forming galaxies \citep{use15}. From left to right: CO, HCN, and \HCO.
\label{fig:sfr-L_global}}
\end{figure*}

\section{DISCUSSION}

\subsection{Scatter in Schmidt-Kennicutt relations \label{sec:discussion}}

The critical densities for HCN and \HCO~(up to $n_{crit}$ $\sim$ 10$^{5-6}$ cm$^{-3}$) to be excited are much higher compared to that of CO ($n_{crit}$ $\sim$ 10$^{3}$ cm$^{-3}$), and therefore they are expected to trace the dense gas that is more relevant to star formation. In this regard, it is natural to expect a tighter correlation in the dense gas-based SK relation than in the CO-based SK relation, as illustrated by early works \citep[e.g.,][]{gao04,wu05,lad10}. This trend, however, is not obvious in our data, in which we show that the correlations between SFR and gas surface density, when limited to the star-forming spaxels, are in fact comparable with each other when using different molecular gas tracers (see Figure \ref{fig:sfr-M_all}). This might imply that for the two main sequence galaxies considered here, either HCN or \HCO~actually traces gas with density lower than their critical densities \citep{har19,eva20}, for example, through the radiative trapping effects or because of the contributions of high masses of the low-density gas \citep{shi15,ler17,jim19}, or they are excited through other mechanisms, such as electron excitation \citep{gol17}, UV from the young massive stars, X-ray from AGNs, and cosmic rays from supernova explosions \citep{pap14,top16}, which influence the molecular line ratios in different ways \citep{mei07,top16}.

Before seeking the physical explanations, we first consider whether our results are impacted by the lower signal-to-noise ratios (S/N) of HCN and \HCO~ lines when compared to that of CO. This discrepancy in S/N ratios may potentially lead to weaker correlations in the dense gas SK relation compared to the CO-based SK relation. To examine this effect, we repeat the correlation analyses for the star-forming spaxels in main sequence galaxies but impose an upper limit of S/N(CO) = 10 this time. The Pearson correlation coefficient of the CO-based SK relation is found to reduce from 0.8 to 0.6, becoming comparable to the values for HCN (0.5) and \HCO (0.7). This test suggests that whether or not the CO-based SK relation is superior to the dense gas SK relation is highly sensitive to the S/N ratios achieved by a given molecular line.

In addition to the effect from the S/N ratio, we note that the dynamical range in the dense gas surface density in our sample is rather narrow  and the number of data points is small, compared to other studies (e.g., EMPIRE shown as magenta in Figure \ref{fig:sfr-M_all}). These might also impact the measurements of intrinsic of correlations.

Next, we consider the possibility where HCN or \HCO~are excited through processes other than collisional excitation. In this case, the HCN/\HCO~ratio can vary depending on which mechanism plays a dominant role. Numerical models predict that the flux ratio of HCN/\HCO~can be greater than unity in photon dissociation regions whereas HCN/\HCO~$<$ 1 in X-ray dissociation regions if the density exceeds 10$^{5}$ cm$^{-3}$ and if the column density is larger than 10$^{23}$ cm$^{-2}$\citep{mei07,top16}. On the other hand, XDR HCN/\HCO~ ratio becomes larger than one if the column density is less than 10$^{22.5}$ cm$^{-2}$ \citep{mei07}. While it is not possible to robustly constrain column densities with current datasets, it is nevertheless insightful to look at the HCN/\HCO~ratio in our sample. 

Figure \ref{fig:his_fhcnhco} shows the histograms of the HCN/\HCO~ratio for the 5 galaxies individually.  One can see that 4 out of 5 galaxies have HCN/\HCO~ $>$ 1 in most spaxels. The BPT diagnostics show no signs of AGN spaxels in these 4 galaxies, suggesting that the physical condition of the interstellar medium (ISM) is consistent with photodissociation regions (PDRs). On the other hand, galaxy 8081-12703, one of the GV galaxies, has the spaxel distributions centered at HCN/\HCO~ratio $\sim$ 1, systematically lower than the values of other galaxies. About half the spaxels have the HCN/\HCO~ratio less than 1, indicating that this galaxy might be affected by either AGN or supernova explosions. The BPT diagnostics using the \othree/\hb~vs. \ntwo/\ha~ (\othree/\hb~vs. \ntwo/\ha) ratios reveal that the central spaxels are consistent with composite (star-forming + LINER regions) and may indicate the presence of central AGN.  Nevertheless, this galaxy only gas very few star-forming spaxels (Figure \ref{fig:sfr-M}) that contribute to the dense gas SK relation of star-forming spaxels (Figure \ref{fig:sfr-M_all}). 

Based on the above tests and evidences, we therefore conclude that the `tighter' relation in the CO-based SK relation compared to the dense gas SK relation we see in Figure \ref{fig:sfr-M_all} is more likely driven by the difference in the S/N ratios and also impacted by the small number statistics rather than a genuine behavior. While the lower S/N ratios of the dense gas tracers compared to CO result in a small dynamical range of \sigdense~ that impacts the fits of the SK relation, our finding that the spaxels in the green valley (GV) exhibit lower \SFEdense~ compared to those in the star-forming main sequence (MS) remains valid. This difference is nearly an order of magnitude at a given \sigdense, significantly larger than the uncertainties (less than 40\%) associated with the S/N ratio cuts.

Finally, we note that the wide range of \SFEdense~found in this work, spanning two orders of magnitude, seems to be in contrast to the constant \SFEdense~model, in which \SFEdense~is approximately constant once the gas surface density is above a certain threshold \citep{lad12,eva14}. On the other hand, our results are consistent with earlier observational studies on kiloparsec (kpc) scales, which have also shown that the dense gas star formation efficiency (\SFEdense) varies with galactic environments and galactocentric radii \citep[e.g.,][]{jim19,neu23}. Therefore, our findings align better with turbulence-regulated models than with fixed-density models \citep{che15,use15,big16,gal18,que19,jim19}.

One of the keys to study the role of turbulence in shaping SFE is through the study of the velocity dispersion ($\sigma$) of the molecular gas at scales close to the moelcular clouds. Previous high-resolution observations ($\sim$ 100pc) revealed an strong anti-correlation between \SFEdense~ and $\sigma$ \citep[e.g.][]{use15,ler17,que19} as model predictions \citep{mei20} while some others found the dependence of \SFEdense~on $\sigma$ is only moderate \citep{san22}. Owing to the coarse spatial resolution and moderate S/N of the dense gas observations used in this work, it is not feasible to directly examine the link between SFE and velocity dispersion. Future higher resolution data with better sensitivity will further shade light on the physical conditions setting \SFEdense~in green valley galaxies.

\subsection{Comparison with previous studies in the literature \label{sec:literature}}

\begin{figure*}
\centering
\includegraphics[angle=0,width=0.95\textwidth]{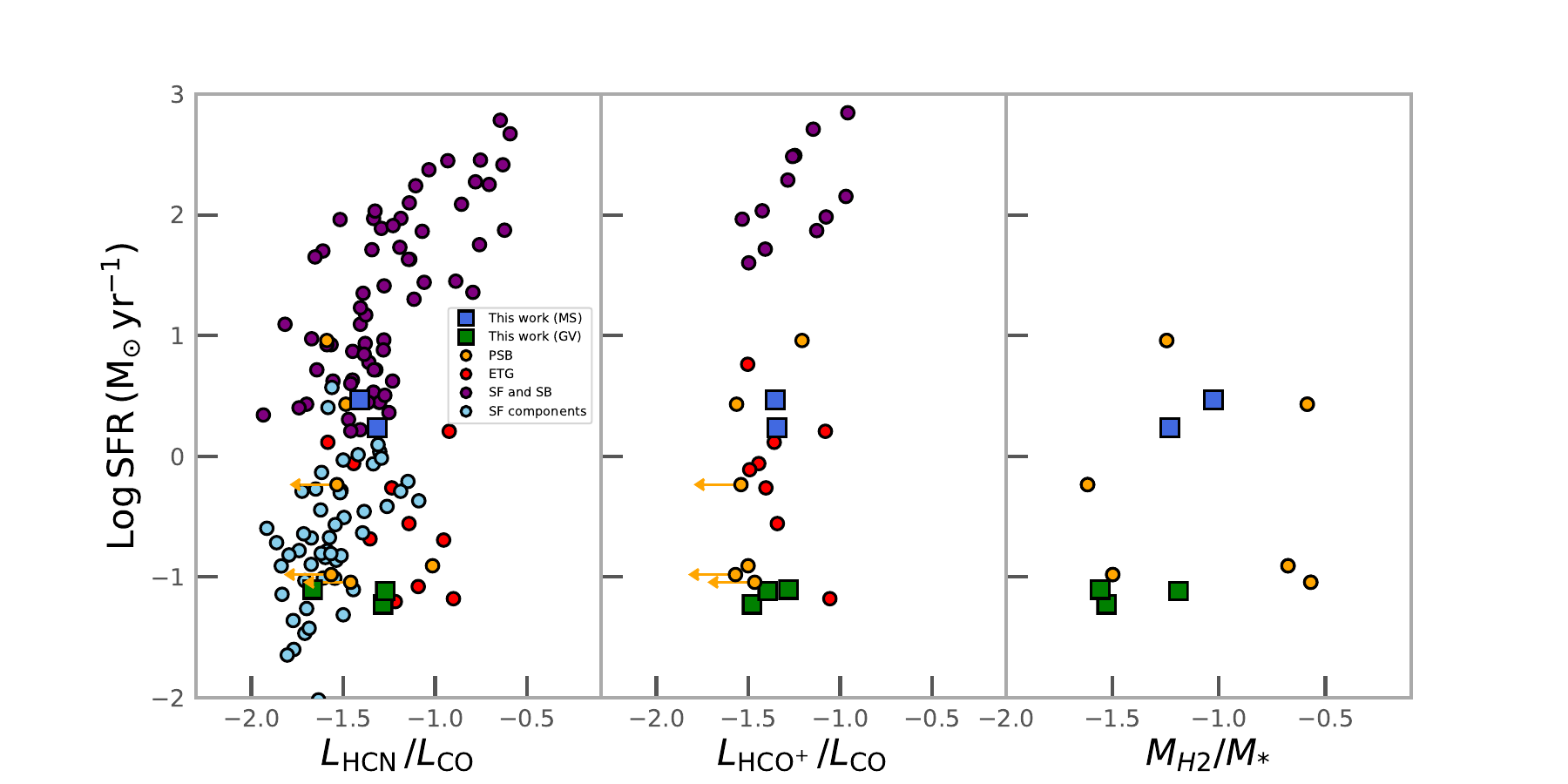}

\caption{The (global) SFR as a function of dense gas luminosity ratio using $L_{\rm{HCN}}$/$L_{\rm{CO}}$ (left panel), $L_{\rm{HCO^{+}}}$/$L_{\rm{CO}}$ (middle panel), and H2-to-stellar mass ratio (right panel) for ALMaQUEST galaxies used in this work (main-sequence: blue squares; green valley: green squares), post-starburst (PSB) galaxies \citep{fre23}, ETGs \citep{cro12}, star-forming and starburst galaxies \citep{gao04}, and subcomponents of star-forming galaxies \citep{use15}.
\label{fig:sfr-fgas_global}}
\end{figure*}

After examining the resolved dense gas properties in our sample, we now compare the global dense gas properties of the 5 ALMaQUEST sources with other galaxy populations to better understand the role of dense gas in the broader context. Figure \ref{fig:sfr-L_global} show the global SFR versus global molecular gas luminosity ($L$) for the 5 ALMaQUEST galaxies and various samples from the literature, including EMPIRE galaxies \citep{jim19}, post-starburst galaxies \citep{row15,fre18,fre23}, early-type galaxies \citep{cro12}, star-forming and starbursting galaxies \citep{gao04}, and subregions in star-forming galaxies \citep{use15}. Most of the star-forming populations, including the star-forming galaxies and starbursting galaxies, follow a linear relation for all gas tracers (CO, HCN, and \HCO) spanning over three orders of magnitude. In contrast, the post-starburst galaxies and ETGs show diverse behaviors, with some on the same trend formed by star-forming galaxies while some others offset from the main relationship toward lower \SFEmol~ as well as lower \SFEdense. The 3 ALMaQUEST green valley galaxies are all below the star-forming linear relationship, similar to those post-starburst galaxies and ETGs that show low \SFEmol~and \SFEdense. These measurements suggest that there is quite a diversity in these transitioning and/or quenched galaxies (including green vallley, post-starburst, and early-type galaxies) in terms of their SFE. Some may have normal SFE as typical star-forming or even starburst galaxies whereas some others have relatively low (dense gas) SFE.

In figure \ref{fig:sfr-fgas_global}, we plot the global SFR versus the ratio of dense molecular gas to total molecular gas traced by $L_{\rm{HCN}}$/$L_{\rm{CO}}$ (left panel) and $L_{\rm{HCO^{+}}}$/$L_{\rm{CO}}$ (right panel) for various types of galaxies. Here we use the line luminosity ratio instead of the gas mass ratio as the dense gas fraction indicator in order to compare different samples in the literature in a more straightforward way. First thing to notice is that at a given population, there is a wide spread in both the $L_{\rm{HCN}}$/$L_{\rm{CO}}$ and $L_{\rm{HCO^{+}}}$/$L_{\rm{CO}}$ ratios. There seems to be no strong systematic difference in the ratio of dense molecular gas to total molecular gas with respect to the level of SFR, except that the starburst sample with the highest SFR does extend to the high end of the dense gas fraction. This implies that the dense gas fraction may not play a dominant role in separating the transitioning or quiescent galaxies from the star-forming population. Among the non-star-forming categories, our green valley sample has comparable dense gas luminosity ratio as ETGs and at least half of the post-starburst galaxies that are detected in dense gas tracers. On the other hand, the post-starbursts show a wider distribution in the gas-to-stellar mass ratio compared to the green valley galaxies targeted in this work as revealed in the right panel of Figure \ref{fig:sfr-fgas_global}.  The 3 PSBs with highest gas-to-stellar ratio are not necessarily the ones associated with the highest dense gas fractions. There is clear diversity in the gas content of the post-starburst galaxies. 
In the future, it will be insightful to look into the connection between the dense gas fraction and the star formation histories with a larger sample to understand better whether there is evolution of the dense gas state along different star formation phases.

\section{SUMMARY \label{sec:summary}}

In this work, we present ALMA HCN(1-0) and \HCO(1-0) observations for 5 galaxies selected from the ALMaQUEST survey, including 3 green valley (GV) galaxies and 2 main-sequence (MS) galaxies. These galaxies are previously selected to exhibit comparable molecular gas fractions (\fmol) traced by their CO abundance but differences in their molecular gas star formation efficiency (\SFEmol), resulting in the difference in the specific star formation rate (sSFR). The aim of this paper is to investigate if the variation in \SFEmol~between MS and GV galaxies is driven by the lack of dense molecular gas, despite the normal molecular gas abundance, or is due to the low efficiency of forming stars with respect to the dense gas abundance, characterized by \SFEdense. Our main findings can be summarized as below:\\

1. Both HCN and \HCO~are detected in all the five galaxies. The global HCN-to-CO and \HCO-to-CO line ratios are found to be 0.02-0.05 and 0.03-0.05, respectively, broadly consistent with the values found in the literature.\\

2. The dense gas Schmidt-Kennicutt relation is found to be different between MS and GV galaxies, with the latter showing lower \sigsfr~for a given \sigdense~ and hence lower \SFEdense~(Figure \ref{fig:sfr-M} and Figure \ref{fig:sfr-M_all}). When separating the star-forming and non-star-forming spaxels, it is seen that the non-star-forming spaxels, which dominate the regions of GV galaxies, deviate significantly from the dense gas SK relation formed by the star-forming spaxels toward lower values of \sigsfr~at a given dense gas surface density.\\

3. The molecular gas based \SFEmol~ (traced by CO) better correlates with \SFEdense~(traced by HCN or \HCO) than with the dense gas fraction, i.e., dense-to-molecular gas mass ratio (\fdense) (Figure \ref{fig:sfe-fgas} and Figure \ref{fig:sfe-sfe}). In fact, the ranges of the \fdense~ratio between star-forming and non-star-forming spaxels, or between MS and GV spaxels, are similar. In other words, \SFEmol~is primarily set by the ability of dense gas to form stars in our sample.\\

4. When looking at the dependence of sSFR on various parameters, we find that sSFR is best correlated with the \SFEmol~and \SFEdense~ (with correlation coefficients $\sim$ 0.8-0.9), followed by the dependence on the molecular gas fraction (\fmol), and is least associated with \fdense~ or the dense-to-stellar mass ratio (Figures \ref{fig:ssfr-sfe}, \ref{fig:ssfr-gastosm}, and \ref{fig:ssfr-gas}). The correlation between sSFR and SFE extends to non-star-forming spaxels as well. Our results suggest that the quenching mechanisms responsible for lowering the star formation activities in these galaxies mainly impact \fgas~and \SFEdense, and do not alter the dense gas-to-molecular gas ratio directly.\\

5. The ALMaQUEST green valley (GV) galaxies exhibit systematically lower global \SFEmol~and \SFEdense~when compared to star-forming and starburst galaxies in previous studies (Figure \ref{fig:sfr-L_global}). However, there is no significant distinction between GV galaxies and other types of galaxies in terms of the dense gas luminosity ratio (Figure \ref{fig:sfr-fgas_global}). This aligns with the spatially resolved findings that the low sSFR of our GV galaxies is not attributed to a lack of dense gas.

In conclusion, we find that the 3 GV galaxies in our sample, while having comparable molecular gas fraction (\fgas), their spaxel-based \fdense~ are also comparable to that in MS galaxies whereas the \SFEmol~ and \SFEdense~ are systematically lower than those of MS galaxies. In other words, these GV galaxies are less efficient in converting dense gas to stars and therefore lead to lower specific star formation rates as opposed to MS galaxies. Nevertheless, We emphasize that the 3 GV galaxies in our sample are CO-abundant by selection and may not be representative of the whole GV population in general. Having a larger GV sample covering a different parameter space may shed light on the connection between star formation and molecular gas, as well as on the physical processes that suppress star formation in GV galaxies.

\acknowledgments
We thank the anonymous referee for his/her helpful comments, which improve the contents of this paper. 
This work is supported by the Ministry of Science \& Technology of Taiwan under the grants MOST 111-2112-M-001-044 - and NSTC 112-2112-M-001-062 -. HAP acknowledges support by the National Science and Technology Council of Taiwan under grant 110-2112-M-032-020-MY3. N.H. acknowledges support from JSPS KAKENHI grant No. JP21K03634. S.F.S. thanks the support by the PAPIIT-DGAPA AG100622 and CONACYT CF19-39578 projects.


The authors would like to thank the staff of the East-Asia and North-America ALMA ARCs for their support and continuous efforts in helping produce high-quality data products. This paper makes use of the following ALMA data:\\
ADS/JAO.ALMA\#2015.1.01225.S,\\  ADS/JAO.ALMA\#2017.1.01093.S,\\
ADS/JAO.ALMA\#2018.1.00541.S,\\
ADS/JAO.ALMA\#2018.1.00558.S.\\
ADS/JAO.ALMA\#2019.1.01178.S.\\
ALMA is a partnership of ESO (representing its member states), NSF (USA) and NINS (Japan), together with NRC (Canada), MOST and ASIAA (Taiwan), and KASI (Republic of Korea), in cooperation with the Republic of Chile. The Joint ALMA Observatory is operated by ESO, AUI/NRAO and NAOJ.

Funding for the Sloan Digital Sky Survey IV has been
provided by the Alfred P. Sloan Foundation, the U.S.
Department of Energy Office of Science, and the Participating Institutions. SDSS-IV acknowledges support
and resources from the Center for High-Performance
Computing at the University of Utah. The SDSS web
site is www.sdss.org. SDSS-IV is managed by the Astrophysical Research Consortium for the Participating
Institutions of the SDSS Collaboration including the
Brazilian Participation Group, the Carnegie Institution
for Science, Carnegie Mellon University, the Chilean
Participation Group, the French Participation Group,
Harvard-Smithsonian Center for Astrophysics, Instituto
de Astrof\'isica de Canarias, The Johns Hopkins University, Kavli Institute for the Physics and Mathematics of the Universe (IPMU) / University of Tokyo, Lawrence
Berkeley National Laboratory, Leibniz Institut f\"ur Astrophysik Potsdam (AIP), Max-Planck-Institut f\"ur Astronomie (MPIA Heidelberg), Max-Planck-Institut f\"ur
Astrophysik (MPA Garching), Max-Planck-Institut f\"ur
Extraterrestrische Physik (MPE), National Astronomical Observatory of China, New Mexico State University,
New York University, University of Notre Dame, Observat\'ario Nacional / MCTI, The Ohio State University,
Pennsylvania State University, Shanghai Astronomical
Observatory, United Kingdom Participation Group, Universidad Nacional Aut\'onoma de M\'exico, University of
Arizona, University of Colorado Boulder, University of
Oxford, University of Portsmouth, University of Utah,
University of Virginia, University of Washington, University of Wisconsin, Vanderbilt University, and Yale University.

\end{document}